\pdfoutput=1
\documentclass{aa}   %% Astronomy & Astrophysics style class v8.2

\usepackage{graphicx,natbib,url,twoopt}
\usepackage[varg]{txfonts}           %% A&A font choice
\usepackage{hyperref}   
\usepackage{booktabs}
\usepackage{url}            %% for pdflatex
\usepackage{pdfcomment}              %% for popup acronym meanings
\usepackage{acronym} 
\usepackage{epstopdf}                %% for popup acronym meanings
\usepackage{bm}
\usepackage{longtable}
\usepackage{caption}
\usepackage{mathtools}
\usepackage{bm}
\usepackage{esvect}
\usepackage{amsmath}
\usepackage{multirow}
\usepackage{adjustbox}
\usepackage{xcolor}
\usepackage{physics}

\hypersetup{
  colorlinks=true,   %% links colored instead of frames
  urlcolor=blue,     %% external hyperlinks
  linkcolor=red,     %% internal latex links (eg Fig)
}

\bibpunct{(}{)}{;}{a}{}{,}    %% natbib cite format used by A&A and ApJ

\makeatletter
\newcommand{\bibnote}[2]{\global\@namedef{#1note}{#2}}
\newcommand{\biblink}[2]{\global\@namedef{#1link}{#2}}

\newcommand{\T}{$T_{\text{X}}$}
\newcommand{\Mgas}{$M_{\text{gas}}$}
\newcommand{\Lx}{$L_{\text{X}}$}
\newcommand{\Yx}{$Y_{\text{X}}$}

\makeatother

\makeatletter
 \newcommandtwoopt{\citeads}[3][][]{%
   \nonstopmode%              %% fix to not stop at error message in latex
   \href{http://adsabs.harvard.edu/abs/#3}%
        {\def\hyper@linkstart##1##2{}%
         \let\hyper@linkend\@empty\citealp[#1][#2]{#3}}%   %% Rutten, 2000
   \biblink{#3}{\href{http://adsabs.harvard.edu/abs/#3}{ADS}}%
   \errorstopmode}            %% fix to resume stopping at error messages 
 \newcommandtwoopt{\citepads}[3][][]{%
   \nonstopmode%              %% fix to not stop at error message in latex
   \href{http://adsabs.harvard.edu/abs/#3}%
        {\def\hyper@linkstart##1##2{}%
         \let\hyper@linkend\@empty\citep[#1][#2]{#3}}%     %% (Rutten 2000)
   \biblink{#3}{\href{http://adsabs.harvard.edu/abs/#3}{ADS}}%
   \errorstopmode}            %% fix to resume stopping at error messages
 \newcommandtwoopt{\citetads}[3][][]{%
   \nonstopmode%              %% fix to not stop at error message in latex
   \href{http://adsabs.harvard.edu/abs/#3}%
        {\def\hyper@linkstart##1##2{}%
         \let\hyper@linkend\@empty\citet[#1][#2]{#3}}%     %% Rutten (2000)
   \biblink{#3}{\href{http://adsabs.harvard.edu/abs/#3}{ADS}}%
   \errorstopmode}            %% fix to resume stopping at error messages 
 \newcommandtwoopt{\citeyearads}[3][][]{%
   \nonstopmode%              %% fix to not stop at error message in latex
   \href{http://adsabs.harvard.edu/abs/#3}%
        {\def\hyper@linkstart##1##2{}%
         \let\hyper@linkend\@empty\citeyear[#1][#2]{#3}}%  %% 2000
   \biblink{#3}{\href{http://adsabs.harvard.edu/abs/#3}{ADS}}%
   \errorstopmode}            %% fix to resume stopping at error messages 
\makeatother

\newacro{ADS}{Astrophysics Data System}
\newacro{NLTE}{non-local thermodynamic equilibrium}
\newacro{NASA}{National Aeronautics and Space Administration}

\begin{document}  

\title{X-ray and optical analysis of the distant, merging double cluster SPT-CLJ2228-5828, its gas bridge, and shock front}

\author{K. Migkas$^{1,2,3}$, M. W. Sommer$^2$, T. Schrabback$^{4,2}$, E. R. Carrasco$^5$, A. Zenteno$^6$, H. Zohren$^2$, L. E. Bleem$^{7,8}$,  V. Nazaretyan$^2$, M. Bayliss$^9$, E. Bulbul$^{10}$, B. Floyd$^{11}$, R. Gassis$^9$, M. McDonald$^{12}$, S. Grandis$^4$, C. Reichardt$^{13}$, A. Sarkar$^{12}$, K. Sharon$^{14}$, T. Somboonpanyakul$^{15}$}

\institute{$^1$ Leiden Observatory, Leiden University, PO Box 9513, 2300 RA Leiden, the Netherlands  \\ 
\email{kmigkas@strw.leidenuniv.nl}\\
$^2$Argelander-Institut f{\"u}r Astronomie, Universit{\"a}t Bonn, Auf dem H{\"u}gel 71, 53121 Bonn, Germany\\
$^3$SRON Netherlands Institute for Space Research, Niels Bohrweg 4, NL-2333 CA Leiden, the Netherlands\\
$^4$Universit{\"a}t Innsbruck, Institut f{\"u}r Astro- und Teilchenphysik, Technikerstr. 25/8, 6020 Innsbruck, Austria \\
$^5$International Gemini Observatory/NSF NOIRLab, Casilla 603, La Serena, Chile.\\ 
$^6$Cerro Tololo Inter-American Observatory/NSF NOIRLab, Casilla 603, La Serena, Chile.\\
$^7$High Energy Physics Division, Argonne National Laboratory, 9700 South Cass Avenue, Lemont, IL 60439, USA\\
$^8$Kavli Institute for Cosmological Physics, University of Chicago, 5640 South Ellis Avenue, Chicago, IL, 60637, USA\\
$^9$Department of Physics, University of Cincinnati, Cincinnati, OH 45221, USA\\
$^{10}$Max Planck Institute for Extraterrestrial Physics, Gie\ss enbachstra\ss e 1, 85748 Garching bei M{\"u}nchen, Germany\\
$^{11}$Institute of Cosmology and Gravitation, University of Portsmouth, Burnaby Road, Portsmouth, PO1 3FX, UK\\
$^{12}$ Kavli Institute for Astrophysics and Space Research, Massachusetts Institute of Technology, 70 Vassar Street, Cambridge, MA 02139, USA\\
$^{13}$ School of Physics, University of Melbourne, Parkville, VIC 3010, Australia\\
$^{14}$Department of Astronomy, University of Michigan, 1085 S. University Ave, Ann Arbor, MI 48109, USA\\
$^{15}$Department of Physics, Faculty of Science, Chulalongkorn University 254 Phayathai Road, Pathumwan, Bangkok Thailand, 10330\\
}

\date{Received date} 
\abstract{Galaxy cluster mergers are excellent laboratories for studying a wide variety of different physical phenomena. Such a unique cluster system is the distant SPT-CLJ2228-5828 merger located at $z\approx 0.77$. Previous analyses via the thermal Sunyaev-Zeldovich effect and weak lensing data suggested that the system potentially was a dissociative cluster post-merger, similar to the Bullet cluster. In this work, we perform an X-ray and optical follow-up analysis of this rare system. We use new, deep \textit{XMM-Newton} data to study the hot gas in X-rays with great detail, spectroscopic \textit{Gemini} data to precisely determine the redshift of the two mass concentrations, and new HST data to improve the total mass estimates of the two components. We find that SPT-CLJ2228-5828 constitutes a pre-merging, double cluster system, instead of a post-merger as previously thought. The merging process of the two clusters has started with their outskirt gas colliding with a $\sim 22^{\circ}-27^{\circ}$ on the plane of the sky. Both clusters have a similar radius of $R_{500}\sim 700$ kpc with the two X-ray emission peaks separated by $\approx 1$ Mpc ($2.1^{\prime}$). We fully characterize the surface brightness, gas density, temperature, pressure, and entropy profiles of the two merging clusters for their undisturbed, non-interacting side. The two systems have very similar X-ray properties with a moderate cluster mass of $M_{\text{tot}}\sim (2.1-2.4)\times 10^{14}\ M_{\odot}$ according to X-ray mass proxies. Both clusters show good agreement with known X-ray scaling relations when their merging side is ignored. The weak lensing mass estimate of the Western cluster agrees well with the X-ray-based mass while, surprisingly, the Eastern cluster is only marginally detected from its weak lensing signal. A gas bridge with $\approx 333$ kpc length connecting the two merging halos is detected at a $5.8\sigma$ level. The baryon overdensity of the excess gas (not associated with the cluster gas) is $\delta_{\text{b}}\sim (75-320)$ across the length of the bridge and its gas mass is $M_{\text{gas}}\sim 1.4\times 10^{12}\ M_{\odot}$. Gas density and temperature jumps at $\sim 10^{-3}$ cm$^{-3}$ and $\sim 5.5$ keV, respectively, are also found across the gas bridge, revealing the existence of a weak shock front with a Mach number $\mathcal{M}\sim 1.1$. The gas pressure and entropy are also increased at the position of the shock front. We estimate the age of the shock front to be $\lesssim 100$ Myr and its kinetic energy $\sim 2.4\times 10^{44}$ erg s$^{-1}$. SPT-CLJ2228-5828 is the first such high-$z$ pre-merger with a gas bridge and a shock front, consisting of similarly-sized clusters, to be studied in X-rays.}
%indicated that the hot intracluster gas might be separated from the two major galaxy and total mass distribution centers. This


\titlerunning{A unique, distant, merging double cluster}
\authorrunning{K. Migkas et al. }

\maketitle

\section{Introduction}\label{intro}

According to the standard hierarchical structure formation model, larger cosmic structures are formed through merging of smaller structures. Galaxy clusters are created at the intersections of the cosmic web through this merging process of smaller substructures. At the later stages of structure formation, nearby galaxy clusters started merging with each other due to their mutual gravitational attraction. Galaxy cluster mergers are the most energetic phenomena in the Universe with colliding velocities of $\sim 1000$ km s$^{-1}$ and released energy of $10^{64}$ ergs \citep{sarazin}. Cluster mergers offer a unique opportunity to study numerous physical processes, ranging from acceleration of particles and cosmic rays to cosmology. For instance, valuable insights on the self-interaction of dark matter particles can be obtained by studying cluster mergers, particularly the offset between the dark matter and gas distributions in post-mergers \citep{Markevitch04,robertson}. Furthermore, the kinetic energy of the clusters can dissipate to the intracluster medium (ICM) and produce shock and cold fronts detected in X-rays by their density and temperature jumps. Such shock and cold fronts can provide information on the microphysics of the ICM \citep[e.g., thermal conduction and magnetic fields,][]{Markevitch03,urmilla22,sarkar23} and its energy transport processes, as well as the dynamics of the merging clusters.

On the other hand, galaxy cluster pre-mergers constitute ideal targets for studying gas bridges that connect individual halos. Analyzing such gas bridges can help us further understand the location of the so-called missing baryons in the late Universe \citep[e.g.,][]{werner}. Shock fronts were also shown to exist in gas bridges between pre-merging clusters \citep[e.g.,][]{kato, sarkar22,omiya} due to colliding gas. Because of the great interest in them, galaxy cluster mergers have been extensively studied in the last decades in the full electromagnetic spectrum and at all merging stages \citep[e.g.,][]{markevitch05,brada06,cassano10,vanWeeren10,tempel17,Ha18}.

Moreover, the study of the individual clusters comprising pre-merging systems can help us determine if they can be used as typical clusters for cosmological and astrophysical studies, such as for constraining the halo mass function and galaxy cluster scaling relations. Interacting clusters are usually disturbed systems, with their level of relaxation (as determined from, e.g., their morphological parameters) depending on their merging stage. Disturbed clusters have been shown to often follow different scaling relations than the general cluster population \citep[e.g.,][]{lovisari20,migkas21} which can bias their use as cosmological probes. However, such studies include the entire systems in their analysis, merging and non-merging cluster sides. No extensive effort has been made to characterize \emph{only} the non-interacting sides of pre-merging systems to determine if their behavior is consistent with isolated, relaxed clusters.

X-ray observations are especially useful for studying cluster mergers. X-ray data from instruments such as \textit{XMM-Newton}, \textit{Chandra}, and \textit{eROSITA}, can spatially resolve the hot gas of merging clusters more efficiently than SZ observations due to their superior angular resolution. X-ray observations also allow for studying the properties of the ICM in much greater detail. Additionally, optical data trace the galaxy population of clusters and their weak and strong lensing signal. Combining X-ray and optical information can provide a complete picture of the state of a merging cluster system and its components. 

In this work, we perform the first X-ray analysis of a previously identified cluster merger, SPT-CLJ2228-5828, alongside a refined weak lensing (WL) analysis of its cluster components. Based on earlier data the cluster was a strong candidate for a post-merger system (see Sect. \ref{cluster-description}). However, we find that SPT-CLJ2228-5828 consists of two individual clusters in a pre-merging phase. The collision of the gas of the two clusters has started, with a gas bridge in-between components where a shock front can be found. While in X-rays the two clusters are highly similar, only one of these clusters shows a clear WL detection, with the other one being marginally detected. 

The structure of the paper is as follows: in Sect. \ref{cluster-description} we present the results of past analyses of the SPT-CLJ2228-5828 merging system. In Sect. \ref{xray-optical-data} we describe the X-ray and optical data used in this work, the data reduction, as well as the imaging, X-ray spectral analysis, and WL analysis. In Sect. \ref{sect:cluster-results}, we present the results regarding the two cluster members of the merger. In Sect. \ref{bridge-sect}, we present the analysis of the gas bridge connecting the two clusters. Finally, in Sect. \ref{sect:summary}, we discuss and summarize our findings. Throughout the paper a $\Lambda$CDM model is assumed with $H_0=70$ km s$^{-1}$ Mpc$^{-1}$ and $\Omega_{\text{m}}=0.3$. 

\section{The SPT-CLJ2228-5828 cluster system}\label{cluster-description}
The SPT-CLJ2228-5828 system was first detected as a single cluster in the 2500 deg$^2$ SPT-SZ survey by \citet{bleem15} at (RA, DEC)$=(337.215^{\circ}, -58.469^{\circ})$ and a photometric redshift $z=0.71\pm 0.06$. From its integrated total Compton parameter the system's total mass within $R_{500}$ was estimated to be $M_{500}=(3.17\pm 0.82)\times 10^{14}$ M$_{\odot}$\footnote{$R_{500}$ ($R_{200}$) refers to the radius of the cluster within which the average cluster density is 500$\times$ (200$\times$) the critical density of the Universe. $M_{500}$ ($M_{200}$) refers to the total mass within this radius.}, pointing to a cluster of average mass. Its apparent radius was determined to be $1.5^{\prime}$, corresponding to $\approx 650$ kpc and classifying it as one of the 43 most disturbed clusters from the 2500 deg$^2$ SPT-SZ survey sample \citep{bleem15}. A follow-up analysis of a subsample of the 2500 deg$^2$ SPT-SZ catalog by \citet{bocquet19} yielded similar results with $M_{500}=3.27^{+0.63}_{-0.83}\times 10^{14}$ M$_{\odot}$ and $z=0.734\pm 0.046$ (photo-$z$), with the system still considered a single cluster. SPT-CLJ2228-5828 was also detected as a single object by \citet{hilton21} using the Atacama Cosmology Telescope and the SZ effect. The determined redshift and mass of the system were $z=0.760\pm 0.013$ (photo-$z$) and $M_{500}=(3.82\pm 0.79)\times 10^{14}$ respectively, consistent with previous studies. \citet{zenteno20} performed a joint SZ and optical analysis of the system which revealed two distinct galaxy overdensities, with the cluster gas, as traced by the SPT-SZ data, located in the middle of these two. The identified brightest cluster galaxy (BCG) of the two galaxy distributions was found to be separated by $0.49\ R_{200}$ from the SZ centroid, corresponding to $1.4^{\prime}$ and $\approx 605$ kpc. Furthermore, \citet{schrabback21b} performed a weak lensing (WL) analysis of the system using shallow \textit{Hubble} Space Telescope (HST) data and found that  there are two peaks in the WL mass reconstruction $\approx 1$ Mpc apart from each other, coinciding with the two galaxy overdensities. 
%The two latter studies strongly indicated that SPT-CLJ2228-5828 was a dissociative cluster post-merger with its gas separated from dark matter, similar to the Bullet cluster \citep{Markevitch04}.
Based on the results of the two latter studies SPT-CLJ2228-5828 is a candidate for a dissociative cluster post-merger with its gas separated from dark matter, similar to the Bullet cluster \citep{Markevitch04}. No X-ray analysis was performed until now on the system since there were no X-ray data. The cluster system was not detected either in the ROSAT All-Sky Survey \citep{rass} nor in the most recent eROSITA All-Sky Survey cluster sample \citep{bulbul24}. In this work, we perform the first X-ray analysis of the SPT-CLJ2228-5828 cluster system using newly obtained deep XMM-Newton data. 

\section{X-ray and optical data}\label{xray-optical-data}

\subsection{XMM-Newton}

The XMM-Newton data used in this work (Obs. ID 0884710101) were obtained  during the observing cycle AO20 with total duration time of 99.1 ks (PI: H. Zohren). This was a dedicated X-ray follow-up observation to better resolve the ICM properties of the system compared to the preexisting SZ analyses.

\subsubsection{Data reduction}
Firstly, we reprocessed the observation data files (ODF) with the HEASoft 6.29 and XMMSAS v19.1.0 software packages. We used the \texttt{emchain} and \texttt{epchain} tasks to
generate calibrated event lists. We selected only good events from the calibrated event lists, excluding bad pixels. The data were then treated for solar flare contamination. We created individual binned light curves in the 0.3-10 keV band from the entire Field Of View (FOV) for all three EPIC cameras (MOS1/MOS2/PN). We used 52 s and 26 s bins for MOS and PN cameras respectively. We discarded all time intervals exceeding a $\pm 3\sigma$ Poissonian threshold of the counts bin distribution with respect to the mean. The clean exposure time of the observation is 85.8 ks for MOS1 and MOS2 and 80.1 ks for PN.

To address any residual soft proton contamination, we estimated the IN/OUT ratios \citep{deluca} for each camera, following \citet{migkas24}. In a nutshell, the IN/OUT ratio refers to the surface brightness (SB) within the FOV after masking the central $10^{\prime}$ over the SB of the unexposed corners. Both SB values were measured in the $6-12$ keV band for the MOS detectors and the combined $(5-7) + (10-14)$ keV band for the PN detector. Given the decrease of the effective area beyond $10^{\prime}$ at $>5$ keV energies, the FOV signal is not expected to have significant contributions from sky photons (e.g., cluster photons). The IN/OUT ratio was estimated using both observation data (IN/OUT$_{\text{obs}}$) and filter wheel closed (FWC) data (IN/OUT$_{\text{FWC}}$). By dividing IN/OUT$_{\text{obs}}$ with IN/OUT$_{\text{FWC}}$ we obtained the final IN/OUT$_{\text{final}}$ ratio value. MOS and PN detectors with IN/OUT$_{\text{final}}<1.15$ are generally considered sufficiently filtered from solar flares. For this work's observation, all three EPIC detectors showed IN/OUT$_{\text{final}}<1.04$ which revealed a complete lack of residual solar flare contamination.

Following the methodology of \citet{miriam} and utilizing the count-rate ratios between the $2.5-5$ keV and $0.5-0.7$ keV bands and their linear relation, we also ensured that our X-ray data do not suffer from Solar Wind Charge Exchange contamination \citep[e.g.,][]{carter08}. We then tested the so-called "anomalous state" of the MOS CCDs \citep{kuntz} which revealed that CCD no. 5 of MOS1 is affected by the low energy enhancement associated with the anomalous state, and thus it was discarded from the rest of the analysis. Out-of-time events were obtained and added to the instrumental background (more details below) and eventually subtracted from the data. The masking of point sources and extended sources not related to the cluster merger took place in two steps; firstly, an automatic, wavelet-based source detection algorithm is applied to the co-added MOS1/MOS2/PN (EPIC) image in the 0.5-2 keV band and masked the detected sources of no interest, as described in \citet{pacaud06}. Then a visual inspection of the 0.5-2 keV and 2-10 keV images followed where we manually masked any suspected point sources missed by the wavelet algorithm. The masked regions were then removed from the event lists and were not considered for the spectral analysis. 

\subsubsection{Instrumental background}
XMM-Newton data show a non-negligible level of instrumental, particle-induced background (PIB) that is important to be accounted for properly, especially when low SB systems are studied. The PIB consists of a non-vignetted continuum and fluorescence lines that originate from the interaction between high-energy particles and telescope detectors.

To estimate the PIB of the \textit{XMM-Newton} data we utilize FWC observations which are dominated by the PIB. The FWC data were rescaled to match the count-rates of the individual unexposed EPIC corners ($>925^{\prime \prime}$ from the pointing's center) of the actual observation. The energy bands used for the rescaling of the FWC data were $2.5-5$ keV for both MOS and PN cameras and $8-9$ keV additionally for the MOS cameras. These energy bands are free from fluorescence lines allowing for an easier estimation of the necessary rescaling factor. For each MOS corner and PN quadrant, the PIB was taken from their closest unexposed corners. This procedure allows us to generate PIB images that can be subtracted from the total count images, eventually resulting in PIB-subtracted, count-rate images of the source. Furthermore, PIB spectra for each EPIC detector are produced and utilized in the spectral analysis later on.

Finally, the inner unexposed PN corners show some residual contamination from FOV photons and soft protons, as demonstrated in \citet{marelli}. This can lead to an overestimation of the PN PIB and an underestimation of the spectroscopically measured X-ray temperature ($T_{\text{X}}$). \citet{rossetti} showed that the $10-14$ keV count-rate at $>905^{\prime \prime}$ from the FOV's center can be used to accurately rescale the FWC data and recover the unbiased PIB of PN. Additionally, \citet{migkas24} showed that for PN data with low IN/OUT$_{\text{final}}<1.1$, using the $2.5-5$ keV band at $>925^{\prime \prime}$ from the FOV's center to determine the rescaling factor of the FWC data and the PIB, also returns unbiased results for PN. Thus, the much lower IN/OUT$_{\text{final}}$ values of the data used in this work guarantees a robust estimation of the PN's PIB.

\subsubsection{Imaging analysis and cluster profiles}\label{imaging}
\begin{figure*}[hbtp]
\centering
               \includegraphics[width=1\textwidth]{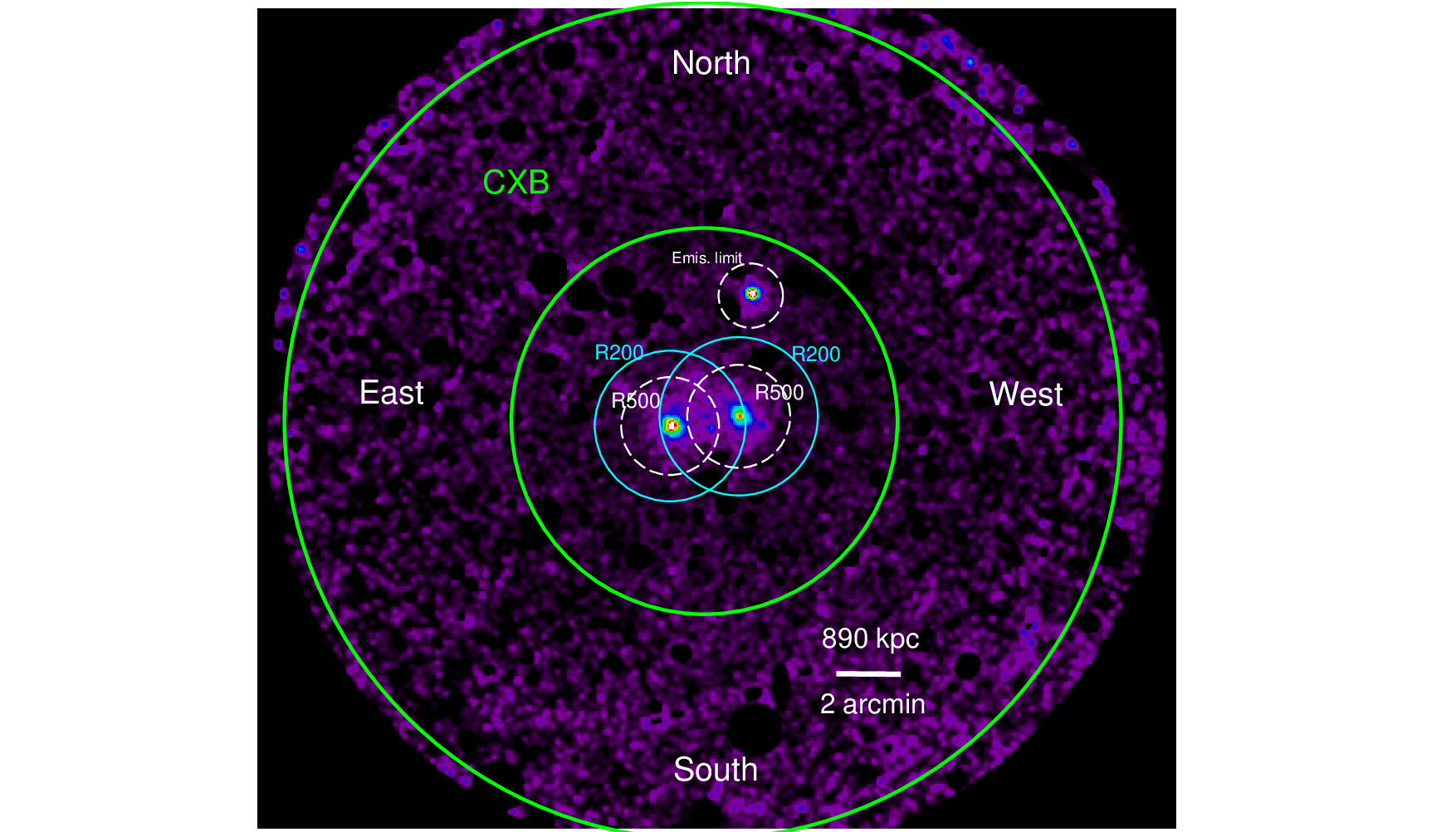}
               \caption{Fully corrected, background-subtracted \textit{XMM-Newton}, smoothed EPIC count-rate image of the SPT-CLJ2228-5828 cluster pre-merger in the $0.5-2$ keV band. The $R_{500}$ and $R_{200}$ radii of each cluster are displayed with white dashed and cyan circles respectively. The region between the solid green circles represents the area of the sky background estimation. The emission limit of the nearby AGN and its host galaxy is also marked with a white dashed circle.}
        \label{XMM-image}
\end{figure*}

To obtain clean count-rate images of the cluster merger we followed all the commonly used steps. Briefly, we firstly subtracted the combined EPIC PIB image from the respective combined source counts image. This resulted in $\approx 3,900$ EPIC clean source counts for each cluster in the $0.5-2$ keV band. Then, we divided the produced image with the effective exposure map which accounts for the absolute exposure time and the vignetting of the detectors due to the decrease of the effective area with the off-axis angle. The effective exposure map additionally accounts for the different effective areas of the PN and MOS cameras for each given energy band. This results to the final, absorbed, combined EPIC count-rate image. To acquire the respective unabsorbed image, we correct the former for the X-ray Galactic absorption using the afore-mentioned $N_{\text{tot}}$ value and the energy range of interest. We check if $N_{\text{tot}}$ significantly varies within the \textit{XMM-Newton} FOV but we do not find any significant changes. The final, fully corrected, smoothed X-ray image of the cluster system in the $0.5-2$ keV band is shown in Fig. \ref{XMM-image}.

To extract and fit the SB and gas density (which is meant to be the electron number density $n_{\text{e}}$ throughout the paper if not stated otherwise) profiles we used the \texttt{pyproffit} package \citep{eckert20} which properly deconvolves the X-ray cluster profiles for the limited point spread function (PSF) of \textit{XMM-Newton}. By integrating the SB profile over the full cluster area one computes the X-ray flux ($f_{\text{X}}$) and luminosity of a cluster ($L_{\text{X}}$). \texttt{pyproffit} also applies a multiscale decomposition of the profiles to compute the 3D, deprojected $n_{\text{e}}$ profile. Integrating the latter over the full cluster volume one obtains the total cluster gas mass $M_{\text{gas}}$. 

To properly study the gas bridge between the two clusters and distinguish it from projection effects due to the superposition of the two cluster outskirts (and signal blending due to the PSF of \textit{XMM-Newton}) we utilized simulated images of the system. Specifically, we firstly simulated two clusters at the positions of the real clusters by adopting the deconvolved surface brightness profiles obtained by \texttt{pyproffit} in the $0.5-2$ kev band. We simulated the cluster emission out to $1.5\ R_{200}$. Moreover, we added a point source (which represents the bright active galactic nuclei, AGN, observed between the two clusters, see Sect. \ref{sect:Sb bridge}) with the same flux as measured by the analysis of the real data. We also uniformly added the CXB level across the full FOV. We did not add any gas bridge emission. Then, we convolved the simulated image with a $15.5^{\prime \prime}$ PSF which accounted for the half-energy width of \textit{XMM-Newton}. Given that the entire cluster system is located close to the FOV center, there are no significant variations of the \textit{XMM-Newton} PSF across the cluster merger. This PSF-convolved simulated image (bottom panel of Fig. \ref{bridge-regions}) illustrates what one would observe with \textit{XMM-Newton} purely due to the emission of the two clusters and their projection effects, assuming no gas bridge. Comparing this simulated image to the real one, we can deduce the existence of excess emission from the assumed gas bridge and infer the bridge's X-ray properties. Finally, to check the robustness of our method, we measured the PSF-convolved surface brightness profiles of the two clusters from the simulated image. We retrieve very similar results to the true observed profiles which highlights the consistency of the simulated and the real image. 

\subsubsection{Spectral analysis}

For the spectral analysis, XSPEC 12.12.0 \citep{xspec} was used. We consider three types of spectra: the cosmic X-ray background (CXB), the PIB, and the source spectra. To extract the CXB spectra, we apply two circular masks of $5^{\prime}$ radius centered in the X-ray peak of each cluster. This radius approximately corresponds to $2\ R_{200}$ for each cluster. We also mask the outer $12^{\prime}-14^{\prime}$ annulus of the XMM-Newton FOV to avoid noisy photon statistics from regions with low effective area. Considering also the point source masks, we are left with $\sim 55\%$ of the entire FOV to constrain the CXB, which is sufficient to provide an accurate and precise characterization of the CXB (see Fig. \ref{XMM-image}). Moreover, the PIB spectra were subtracted from the CXB and source spectra. To account for an imperfect subtraction, we added additional Gaussian lines to the model, representing Al K-$\alpha$ and Si-K-$\alpha$ fluorescence line residuals at $1.485$ keV and $1.7397$ keV respectively. We use the $0.5-7$ keV energy band for the spectral analysis, avoiding the additional PN fluorescence lines in the $7.5-9$ keV band (the high$-z$ of the clusters, and moderate temperature as shown later, also mean there are nearly no detected source photons at $\gtrsim 7$ keV). The full spectral model is
\begin{equation}
\begin{split}
\mathtt{Model =} &\quad\mathtt{constant\times(apec_1 + tbabs\times(apec_2 +}\\
&\quad\mathtt{powerlaw) + tbabs\times apec_3)}.\\
\end{split}
\label{eq:spectral_model}
\end{equation}
The \texttt{constant} (arcmin$^2$) term represents the scaling of the CXB components to the areas of the source regions. The components consist of the unabsorbed thermal emission from the Local Hot Bubble (\texttt{apec$\mathtt{_1}$}) with a temperature free to vary within $T_{\text{X}}\sim 0.08-0.12$ keV\footnote{Throughout the paper we use the notation $k_{\text{B}}T\equiv T_{\text{X}}$, where $k_{\text{B}}$ is the Boltzmann constant and $T$ the temperature measured in Kelvin.} and metallicity $Z\sim 0.9-1.1\ Z_{\odot}$, the absorbed Milky Way Halo emission (\texttt{apec$\mathtt{_2}$}) with  $T\sim 0.18-0.27$ keV and $Z\sim 0.9-1.1\ Z_{\odot}$ \citep{McCammon_2002}, and the cosmic X-ray background from the unresolved point sources (\texttt{powerlaw}) with a photon index of 1.46 \citep[e.g.,][]{Luo_2017}. The Galactic absorption along the line of sight is modeled by a \texttt{tphabs} model and its total hydrogen column density parameter is set to $N_{\text{H,tot}}=1.89\times 10^{20}$ cm$^{-2}$ according to \citet{willingale}. The last term accounts for the source spectra and it is an absorbed thermal emission component (\texttt{tbabs$\times$apec$\mathtt{_3}$}). All properties of each model component (e.g., normalizations, temperatures, and metallicities) were linked across the three EPIC detectors.

Firstly, we fit only the CXB spectra. Then, we use the best-fit values as priors for the CXB model components and we leave the full model free to vary (\texttt{apec1, apec2, powerlaw} and \texttt{apec3}) while fitting the source spectra which are composed of the source$+$CXB photons. Cash statistics (\texttt{cstat} in XSPEC) were used for constraining the best-fit parameters for the model.

In this work, we extract \T\ profiles of the two clusters and the gas bridge with a typical annulus or box width of $\approx 30-32^{\prime \prime}$. Owing to the finite PSF of \textit{XMM-Newton}, photons originating from one region might be falsely detected in another, nearby region. This might lead to an overestimation of \T\ for outer, cooler regions to which photons from hotter regions leaked. To account for this effect, we utilize crosstalk ancillary response files (ARF) between different annuli, created by the SAS task \texttt{arfgen}. These crosstalk ARF account for the contribution of each region to the emission of all the other regions and vice versa. All regions are then fitted simultaneously. This correction effectively removes the photon leakage between different regions due to the limited \textit{XMM-Newton} PSF. 

As a test, we also performed the spectral analysis for each individual cluster annulus individually without applying the crosstalk correction. The \T\ changes are minimal and much smaller than the statistical uncertainties (typically $\lesssim 0.3\sigma$ shifts were found). Thus, even though the crosstalk correction improves the robustness of our analysis, it is not crucial and does not change the scientific conclusions. This was also confirmed by \citet{bartalucci}, who used even narrower cluster annuli than in this work.

\subsection{Hubble Space Telescope observations}

\subsubsection{Data description and reduction}

SPT-CLJ2228-5828 was observed  by the \textit{Hubble} Space Telescope (HST) as part of programme GO 16488 (PI: Hannah Zohren)  using ACS/WFC on March 27th and 28th, 2021, followed by observations obtained with  WFC3/IR  on March 31st, 2021.
All observations were conducted during Continuous Viewing Zone
(CVZ) opportunities which allow for larger total integration times per orbit.
The ACS observations form a $2\times 2$ mosaic, with eight dithered exposures obtained per pointing in the F606W and F814W filters each, resulting in added integration times of 3.57 ks per pointing and filter.
During parts of CVZ orbits HST is observing close to Earth's illuminated limb, leading to increased background levels due to Earthshine. We preferentially scheduled F814W observations during these orbital phases with increased background, while F606W observations are  preferentially obtained
during orbital phases with low background. This maximises the depth of the F606W images, which are used for weak lensing shape measurements (see Sect.\thinspace\ref{se:wl_measurements}). In contrast, the F814W observations are only incorporated into the photometric source selection, which has weaker depth requirements.

The WFC3/IR images were taken in the F140W filter. They consist of 12 partially overlapping exposures that cover the central regions of both cluster components, with a  combined integration time 
of 3.79 ks.

In order to have a homogeneous depth across for the mosaic image area for the WL analysis, we do not incorporate shallower   ACS F606W Snapshot imaging into our analysis, which exist for the central cluster region from programme SNAP 13412 \citep[see][]{schrabback21b}.

The data reduction closely follows \citet{zohren22}, to which we refer the reader for details. In summary, the reduction is largely based on 
the standard ACS (\texttt{CALACS}\footnote{\url{https://hst-docs.stsci.edu/acsdhb}, Chapter 3}) and WFC3 
(\texttt{calwf3}\footnote{\url{https://hst-docs.stsci.edu/wfc3dhb}, Chapter 3})
calibration pipelines, which we complement with 
the algorithm  for the correction of charge-transfer inefficiency in the ACS data described by \citet{massey14}, as well as scripts for refined masking from \citet{schrabback10}. The image alignment and stacking masles use of the \texttt{DrizzlePac}\footnote{\url{https://www.stsci.edu/scientific-community/software/drizzlepac.html}}, as described by \citet{zohren22}.

\subsubsection{Weak lensing measurements}
\label{se:wl_measurements}

Using the  ACS F606W ($V$) images we detect objects using \texttt{SExtractor} \citep{bertin1996} and measure weak lensing (WL) galaxy shapes using the KSB+ formalism 
%\citep{kaiser1995,luppino1997,hoekstra1998}, 
%\citet*{kaiser1995}
(\citealp*{kaiser1995};
\citealp{luppino1997};
\citealp{hoekstra1998})
employing the implementation from \citet{erben2001} and \citet{schrabback07}.
We model the spatially and temporally varying ACS point-spread function employing the  principal component analysis model calibrated by \citet{schrabback10,schrabback18} using stellar fields.
The employed KSB+ pipeline was calibrated on custom ACS-like image simulations by  \citet{hernandez20}, including a correction for multiplicative shear measurement bias that depends on the galaxy signal-to-noise ratio.
For the source selection we incorporate the   ACS F814W ($I$) images, preferentially  selecting background galaxies via magnitude dependent cuts in $V-I$ colour, as detailed in 
\citet{schrabback18,schrabback21b}.
Removing faint sources by requiring $V_{606}<26.5$ AB mag and $S/N_\mathrm{flux}>10$  (define as the ratio of the FLUX\_AUTO and FLUXERR\_AUTO parameters from \texttt{SExtractor}), and applying both shape measurements cuts and the colour selection, our final WL source catalogue comprises 1047 galaxies within the $\sim 6.4^{\prime} \times  6.4^{\prime}$ area covered by the ACS mosaic, corresponding to a density of 25.6 galaxies / arcmin$^2$.
We do not have enough photometric bands to infer reliable photometric redshifts for individual sources in the field. Instead, we follow standard procedures to infer the source redshift distribution from well-studied external references fields, to which statistically consistent source selections are applied.
In particular, our estimation of the source redshift distribution follows \citet{schrabback21b} and is based
on deep data from the CANDELS fields \citep{grogin2011}, employing the recalibrated photometric redshift catalogue from \citet{raihan20} that makes use of photometric measurements from \citet{skelton14}.
This yields a magnitude-dependent estimate for $\langle \beta \rangle$ and $\langle \beta^2 \rangle$, where the lensing efficiency $\beta$ is defined through $\beta=\frac{D_\mathrm{ls}}{D_\mathrm{s}} H(z_\mathrm{s}-z_\mathrm{l})$. Here,  $D_\mathrm{s}$ and $D_\mathrm{ls}$ are the angular diameter distances between the observer and the
source, and the lens and the
source, respectively. The Heaviside step function $H(x)$ is unity for $x > 0$ and zero otherwise.
Weighted according to the magnitude-dependent WL shape weight $w$ their estimated mean values are $\langle \beta \rangle=0.420$ and $\langle \beta^2 \rangle=0.205$ (when assuming our reference cosmological model).

\subsubsection{Mass reconstruction}

\begin{figure*}[hbtp]
               \includegraphics[width=0.99\textwidth]{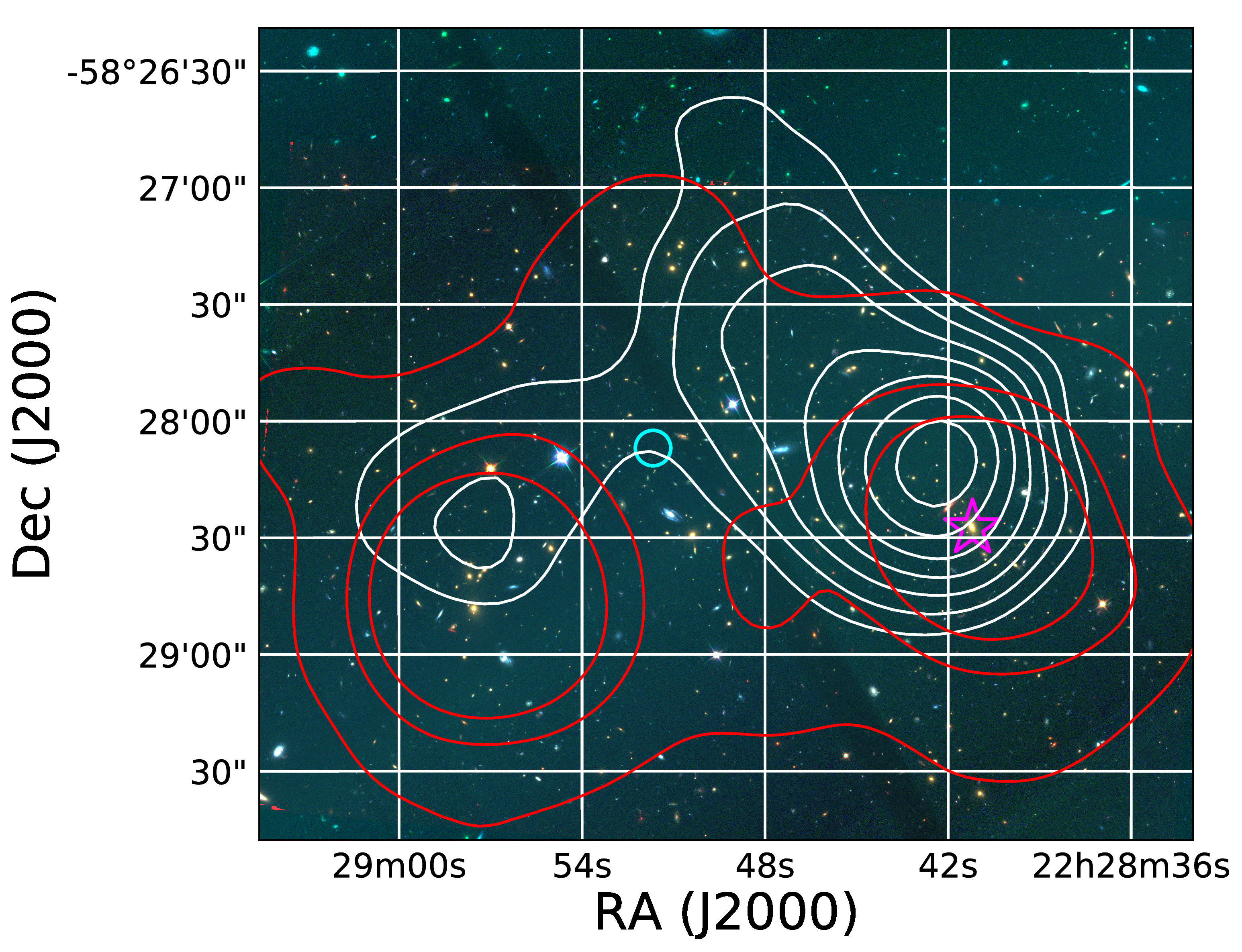}
                \caption{Overlay of the signal-to-noise ratio map of the WL mass reconstruction (shown as white contours starting at level $S/N=1.5$ in steps of  $\Delta S/N=0.5$) and a colour image of the inner cluster regions, employing the HST F140W, F814W, and F606W images for the $R, G, B$ channels. For the non-parametric mass reconstruction shown in this image, the positions of the WL centers were left free to vary.
                %to maximize the signal-to-noise ratio of the WL signal while
                In contrast, for the mass constraints presented in Sect.\thinspace\ref{se:WL_mass_estimation}
                %in the default analysis 
                we fix them to the X-ray cluster centers. The two positions, as well as the BCG positions, agree well within $\lesssim 1\sigma$. The cyan circle indicates the SZ centre from \citet{bleem15}, while the magenta star marks the BCG of the Western cluster component. The red contours show the smoothed X-ray MOS count-rate in steps of $1.385\times 10^{-5}$ cts.
                }
        \label{fig:massmap}
\end{figure*}

%Being both second-order derivatives of the lensing potential  \citep[e.g.][]{bartelmann01} the WL shear $\gamma$ and convergence $\kappa$ are related, allowing for a 
The weak lensing shear $\gamma$ and the convergence $\kappa$ are both second-order derivatives of the lensing potential \citep[e.g.][]{bartelmann01}. This relationship allows for a 
reconstruction of the $\kappa$ distribution from the shear field up to a constant offset, which is often referred to as the  mass-sheet degeneracy \citep[e.g.][]{kaiser93,schneider95}.
%%MS: let us be picky here: the mass sheet degeneracy is formally NOT a constant offset; see review by P. Schneider
We conduct such a reconstruction based on our WL catalogue,  fixing the mass-sheet degeneracy by assuming a mean convergence of zero within the field and employing the algorithm from  \citet{mcinnes09} and \citet{simon09}, which incorporates a Wiener filter.
In order to estimate a signal-to-noise ratio ($S/N$) map of the $\kappa$ reconstruction,
we  compute the root-mean-square (r.m.s.) image  
of  500 noise reconstructions that are based on randomising the ellipticity phases.
The resulting $S/N$ reconstruction is shown in Fig.\thinspace\ref{fig:massmap} overlaid on the HST colour image as white contours. In the image, the Western component of the cluster is detected with peak significance of  $S/N_\mathrm{peak,W}\simeq 4.8$, whereas the Eastern component is  only tentatively detected with   $S/N_\mathrm{peak,E}\simeq 2.3$. 
In the reconstruction, both peaks show a tentative connection at low significance   ($S/N\sim 1.5$--$ 2.0$).
We also note an extension of the Western component to the North-East, coinciding with the location of several potential cluster member galaxies.

\subsubsection{Weak lensing mass estimation}
\label{se:WL_mass_estimation}
The system under study is composed of two interacting clusters of galaxies. To disentangle the masses of the two sub-components from the weak lensing signal, we simultaneously fit for two spherically symmetric Navarro–Frenk–White (NFW) models. The free parameters are the two masses ($M_{200,\rm{~west}}$ and $M_{200,\rm{~east}}$) enclosed within each corresponding $R_{200}$. 
%$M_{200,\rm{~west}}$ and $M_{200,\rm{~east}}$ 
These masses are individually tied to the NFW concentration parameter $c_{200}$ (see below) through the use of the redshift-dependent concentration$-$mass relation of \citet{diemer}, with the corrected parameters of \cite{diemer2}. The centers of the two NFW models were fixed to the coordinates of the X-ray emission peaks.\footnote{Using the BCG coordinates instead has a very small impact upon the mass results.}
%% MS: replaced "\footnote{Using the brightest cluster galaxy coordinates returns almost identical results to the X-ray centers for both clusters.}"

The NFW model parameterizes the mass density $\rho(r)$ at physical radius $r$ by
\begin{equation}
\label{eq:3dnfw}
    \rho(r) = \frac{M_\Delta}{4\pi f(c_\Delta)} \frac{1}{r(r+\frac{r_{\Delta}}{c_{\Delta}})^2},
\end{equation}
where $c_{\Delta}$ is the concentration parameter, and 
$f(c_{\Delta}) = \log(1+c_\Delta) - c_\Delta/(1+c_\Delta)$. Projecting the three-dimensional density model onto the plane of the sky yields the mass surface density
\begin{equation}
\label{eq:projectednfw}
    \Sigma(R) = 2 \int_{0}^{\infty} \rho \left( \sqrt{R^2+\zeta^2}\right) \text{d} \zeta,
\end{equation}
where $R$ is a projected radius and the integration variable $\zeta$ is in the direction of the line of sight. Analytic expressions for the projected surface density $\Sigma$, the convergence $\kappa$ and the shear $\gamma$ of the NFW profile have been provided by \citet{bartelmann96} and \cite{brainerd00}. To model the shear field given two masses and their associated centers, we use the projected surface density (Eq. \ref{eq:projectednfw}) of the NFW model and its associated prediction for the convergence and the tangential shear $\gamma_{\mathrm{t}}$ at the projected radius of each measured ellipticity, excluding any ellipticities with projected radii $R \leq R_{\rm{min}} = 0.2$ Mpc. Note that there are two radii associated with each ellipticity (corresponding to each X-ray or BCG center coordinate), so that effectively two circular regions are excised. We convert the tangential shears of the model into $\gamma_1$ and $\gamma_2$ by the relation
\begin{equation}
\label{eq:shear12}
\begin{array}{lcl}
      \gamma_1 & = & - \gamma_{\mathrm{t}} \cos{(2 \alpha)} + \gamma_{\times} \sin{(2 \alpha)} \\
      \gamma_2 & = & - \gamma_{\mathrm{t}} \sin{(2 \alpha)} - \gamma_{\times} \cos{(2 \alpha)},
\end{array}
\end{equation}
where $\alpha$ is the position angle of a given background galaxy, measured from the respective center, and the cross shear $\gamma_{\times}$ vanishes by construction due to the spherical symmetry of the model. 
%%We apply Eq. \ref{eq:shear12} to each cluster component, and similarly compute the convergence $\kappa$ for both components. 
%% MS: removed a redundant sentence.
%Using individual uncertainties $\sigma_{\bar{g}}$ on the measured ellipticities $\bar{g}_1$ and $\bar{g}_2$, 
%Using magnitude dependent estimates for the uncertainties $\sigma_{\bar{g}}$ on the measured ellipticities $\bar{g}_1$ and $\bar{g}_2$, 
We fit the measured reduced shear estimates provided by the galaxy ellipticities applying magnitude-dependent shear weights \citep{schrabback18} to estimate masses using Markov-Chain-Monte-Carlo sampling. In each step of the chain, we also scale $M_{200}$ to $M_{500}$ to allow a subsequent comparison to the scaling relation of \citet{bulbul19} in Sect. \ref{comparison-scaling}. 
%{\bf Comment Tim: My shear weights are magnitude dependent. I changed the text accordingly. But what did you indicate using the bars?}

Because reduced shear is not additive, we first compute the convergence 
$\kappa_{\infty}$ and shear $\gamma_{\infty}$ at infinite source distance for both components of the model, add the latter, and subsequently compute the total reduced shear of the model as   
\begin{equation}
g = \left( 1 + \left( \frac{\langle\beta_s^2\rangle}{\langle\beta_s\rangle^{2}} -1 \right)  \langle\beta_s\rangle \kappa_{\infty} \right) \frac{\langle\beta_s\rangle \gamma_{\infty}} {1- \langle\beta_s\rangle \kappa_{\infty}},
\end{equation}
which accounts for the $\beta$ correction of \citet{hoekstra}.
Here, $\beta_s = \beta / \beta_{\infty}$, with $\beta_{\infty}$ the lensing efficiency of a source at infinite redshift. 
Our data are the measured ellipticities
$\hat{g}_{k,i}$, where $k \in [1,2]$ and the second subscript indicates the $i$th ellipticity. 
%%\hat{g}_{1,2}$. 
The log-likelihood that we seek to maximize is given by
%%
%% MS: replaced equation, see below
%%
%\begin{equation}
%\begin{split}
%\ln \mathcal{L} &= -\frac{2n}{2}\ln(2\pi) \, - \\
%& \sum_{i=1}^n \left( 2 \ln(\sigma_{\bar{g}_i}) + \frac{1}{2} \left( \frac{\bar{g}_{1,i} - \hat{g}_{1,i}}{\sigma_{\bar{g}_i}}\right)^2 \right. \
%&\left. + \frac{1}{2} \left( \frac{\bar{g}_{2,i} - \hat{g}_{2,i}}{\sigma_{\bar{g}_i}}\right)^2\right),
%\end{split}
%\end{equation}
%%
%% MS: this form of the equation is shorter but perhaps also more confusing (?)
\begin{equation}
\ln \mathcal{L} = -\frac{2n}{2}\ln(2\pi) -
\sum_{i=1}^n 
\left(2 \ln(\sigma_{\hat{g}_i}) + 
\frac{1}{2} \sum_{k=1}^2
\left( \frac{\hat{g}_{k,i} - {g}_{k,i}}{\sigma_{\hat{g}_i}}\right)^2 
%\left( \frac{\bar{g}_{2,i} - \hat{g}_{2,i}}{\sigma_{\bar{g}_i}}\right)^2\right)
\right),
\end{equation}
where $n$ is the number of individual ellipticities measured, and the uncertainties on the reduced shear estimates $\sigma_{\hat{g}_i}=w_i^{-0.5}$ are computed from the shear weights $w_i$ \citep{schrabback18}. 

To determine a 68\% credibility interval of each mass component, we take the shortest interval of the respective marginalized posterior containing 68\% of the samples. The best-fit is taken at the point with the highest density in the marginalized posterior. We test the robustness of the uncertainty estimates $\sigma_{\hat{g}}$ of the ellipticities by subtracting the best-fit model (in terms of $g_1$ and $g_2$) from the data, randomizing the residual ellipticities by rotation, and adding the best-fit model back in. We repeat this 100 times and determine the best-fit masses in each iteration. From the 100 pairs of best-fit masses, we compute the variance for each component. We find that the latter are consistent with the mass variances of the original fit to within 10\%, indicating that the ellipticity uncertainties are robust.  

\subsection{Gemini}

\subsubsection{Observations and data processing}

We selected galaxies for spectroscopy follow-up using catalogs and images\footnote{Downloaded from https://deslabs.ncsa.illinois.edu/desaccess/} from the Dark Energy Survey Year 3 \citep[DESY3][]{sevilla2021}. Galaxies were chosen from those located in the red cluster sequence (RCS) with $(r-i)_{auto}\leq\pm0.22$ from the best fit, and located within then known $R_{200}$ radius \citep[$\leq$ 2.85$^{\prime}$, ][]{zenteno20}. 

To have enough cluster members and to be able to detect any deviation from a Gaussian distribution in the redshift space, about $\sim70-80$ galaxy members are required to be detected for a typical SPT cluster with M$_{200,c} \sim 5 \times 10^{14}$ M$_{\odot}$. Therefore, we have to go as deep as M$^{*}+1$ mag. This magnitude cut-off corresponds to an apparent magnitude of $i \sim 22$ AB mag at the estimated photometric $z$ of SPT-CLJ2228-5828 (no prior spectroscopic $z$ was available). We included 71 galaxies with $i\lesssim22$ AB mag in two designed masks.

Multi-object spectroscopic (MOS) observations were performed using the Gemini Multi-Object Spectrograph \citep[hereafter GMOS,][]{hook04}, mounted at the Gemini South telescope in Chile, in queue mode (program ID: GS-2020B-Q-131). The masks were observed between 23 November 2020 UT and 04 December 2020 UT, during gray time, clear skies, and with a seeing that varied between 0.7$^{\prime \prime}$ and 1$^{\prime \prime}$. The spectra were acquired with the 400 line mm$^{-1}$ ruling density grating (R400), binning by two in both directions, 1$^{\prime \prime}$ slitlets, and central wavelengths of 8600, 8700 and 8800 \AA offsets are applied to avoid loss of any line that could, by chance, end in the gaps between the GMOS CCDs). Due to the faint nature of the galaxies and the central wavelength used, the contamination by the sky lines is expected to be significant. To minimize this effect and obtain reliable redshifts, masks were observed using the \textit{nod-and-shuffle} technique \citep{glaz2001} in \textit{micro shuffling} mode. 

All spectra and corresponding calibrations were processed using the latest version of the Gemini GMOS package inside IRAF, following the standard procedures for \textit{nod-and-shuffle} observations. In summary, science spectra, spectroscopic flats, and comparison lamps are bias subtracted and trimmed. Spectroscopic flats are processed by removing the calibration units' uneven illumination and the corresponding GMOS spectral response, leaving only the pixel-to-pixel variations. Then, the two-dimensional science spectra are flat-fielded, calibrated in wavelength, sky subtracted, and extracted to a one-dimensional format using a fixed aperture of 1.5$^{\prime \prime}$. With the chosen slit width and central wavelengths, the final spectra have a resolution of $\sim 7.3$ \AA, and a dispersion of $\sim 1.34$ \AA\ pixel$^{-1}$. 

\subsubsection{Galaxy redshifts}

We estimated the redshifts of the galaxies by cross-correlating the spectra with high signal-to-noise templates using the routine FXCOR available inside the IRAF RV package. For those spectra with clear emission lines, we used the routine RVIDLINE, employing a line-by-line Gaussian fit to measure the redshifts. We were able to measure the redshift for 59 galaxies out of the 71 selected for spectroscopy ($\sim 83$\% of the sample) in the field of SPT-CLJ2228-5828 ($5.5 \times 5.5$ arcmin$^2$ GMOS field of view). There are ten galaxies for which the signal-to-noise ratio was too low to estimate the redshift correctly, and two galaxies for which the spectra were observed in both masks. %Finally, there are 20 galaxies that lie in the $0.74 \lesssim z \lesssim 0.78$ redshift interval (within $\pm\, 4000$ \kms~ around $z \thickapprox0.7585$). The inset in Fig. \ref{fig:histoall} shows the distribution of the galaxies within $\pm\, 4000$ \kms\ and the locations of the BCG and the second brightest galaxy located in the West and East structures, respectively. 

%\begin{figure}[!h]
%\centering
%\includegraphics[width=0.9\hsize]{SPTCLJ2228_hist.png}
%\caption{Redshift distribution of the 59 galaxies with secure spectroscopic redshifts in the field of SPT-CLJ2228-5828. The  gray region indicates the galaxies at the redshift of the cluster ($z \thickapprox 0.7585$) within $\pm 4000$ \kms ($0.74 \lesssim z \lesssim 0.78$). The tick marks (red for cluster galaxies) at the top of the figure represent the redshift of individual galaxies. The inset shows the peculiar velocity of the galaxies within $\pm\, 4000$ \kms. The vertical lines shows the location of the two central galaxies in the west and east clusters, respectively (see section \S \ref{redshift_sect}).} \label{fig:histoall}
%\end{figure}
%Figure \ref{fig:histoall} shows the redshift distribution for the 59 galaxies in the field of SPT-CLJ2228-5828 observed within the $5.5 \times 5.5$ arcmin$^2$ GMOS field of view. The gray region indicates the galaxies at the redshift of the cluster. 

\section{Analysis of the merging galaxy clusters}\label{sect:cluster-results}

\subsection{Visual inspection and X-ray emission peak}

The clean \textit{XMM-Newton} count-rate image in the $0.5-2$ keV energy band (Fig. \ref{XMM-image}) reveals two separate clusters projected close to each other, only $\approx 2^{\prime}$ apart. Both clusters seem rather symmetrical and relaxed with obvious extended emission beyond their bright centers. From the visual inspection alone, one can determine that this is not a dissociative post-merger system as initially thought from the past SZ-optical and WL analyses, but a probable pre-merger system, or two clusters at different cosmic distances 
projected close to each other on the plane of the sky. Our following analysis confirms that the system is a real pre-merger and gas collision has actually started.

Furthermore, there is an apparent bridge connecting the two clusters that shows excess emission. Although this could be attributed to the superposition of the projected cluster outskirts, in Sect. \ref{bridge-sect} we confirm the existence of the excess gas bridge. North-West of the cluster system, there is a bright, slightly extended source with an apparent size of $\approx 1^{\prime}$ (limit of emission detection compared to CXB, Fig. \ref{XMM-image}). We did not find any registered optical redshift information in the literature for this object. We performed a spectral analysis to the system finding that a \texttt{pow} model fits the central $20^{\prime \prime}$ of the North-West source better than an \texttt{apec} model indicating the presence of a luminous AGN in the center. The $(20^{\prime \prime}-50^{\prime \prime})$ annulus is fitted better with an \texttt{apec} model with $T=0.73^{+0.32}_{-0.25}$ keV and $z=0.08^{+0.09}_{-0.05}$ with an assumed metallicity of $Z=0.3\ Z_{\odot}$. These results suggest that the source is a nearby, hot elliptical galaxy with a $\sim 100$ kpc radius projected onto the plane of the sky.\footnote{A large elliptical galaxy can be indeed seen in the Dark Energy Survey Year 3 \citep{hartley22} at (RA, DEC)$=(337.1598^{\circ}, -58.4109^{\circ})$.} Therefore, we mask it for the rest of the analysis.

We consider the cluster centers to be the X-ray emission peak of each system. To determine the emission peaks, we used the clean count-rate image (produced as described in Sect. \ref{imaging}) in the $0.4-1.25$ keV energy band where the cluster signal maximizes over the CXB. We first smoothed the image by a Gaussian function with a $45^{\prime \prime}$ full width half maximum (FWHM) and the initial X-ray emission peaks were determined. This intentional worsening of the resolution was adopted to ensure that the X-ray emission peaks were not selected based on a poorly removed AGN or a single bright pixel of the image. After this first peak selection, we iteratively improved the Gaussian function FWHM to look for a more precise X-ray emission peak within a circle centered at the first peak with radius five times the previous FWHM. This was done until a final resolution of $7.5^{\prime \prime}$ is reached. The final X-ray cluster centers in equatorial coordinates are (RA$_{\text{Eastern}}$, DEC$_{\text{Eastern}})$=($337.239^{\circ}$, $-58.479^{\circ}$) and (RA$_{\text{Western}}$, DEC$_{\text{Western}})$=($337.171^{\circ}$, $-58.474^{\circ}$) for the Eastern and Western cluster respectively. Their projected angular distance is $2.14^{\prime}$, which corresponds to $951$ kpc according to their $z$ determined in Sect. \ref{redshift_sect}.

\subsection{Redshift determination}\label{redshift_sect}
\begin{figure*}[hbtp]
               \includegraphics[width=0.33\textwidth]{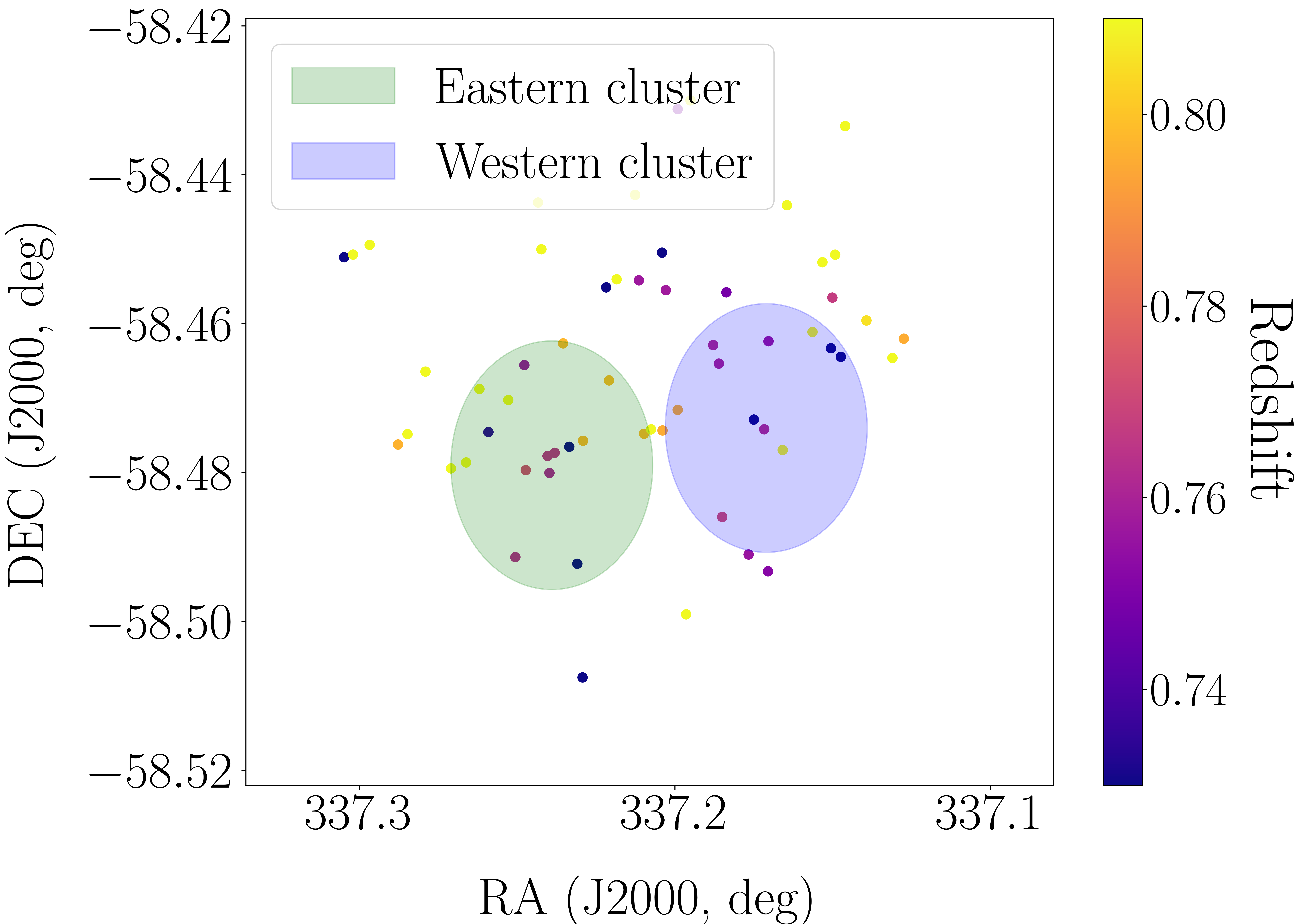}
               \includegraphics[width=0.33\textwidth]{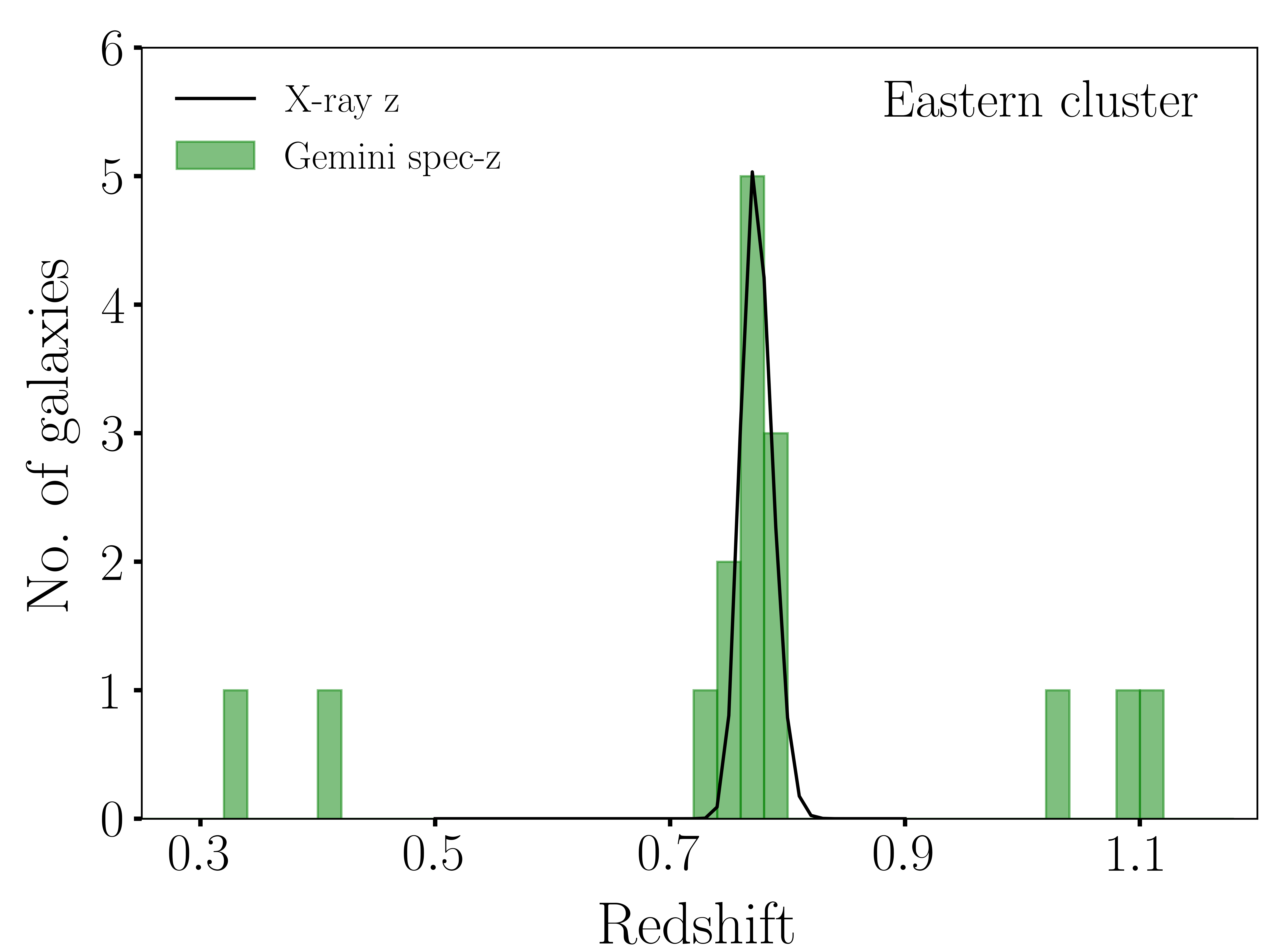}
               \includegraphics[width=0.33\textwidth]{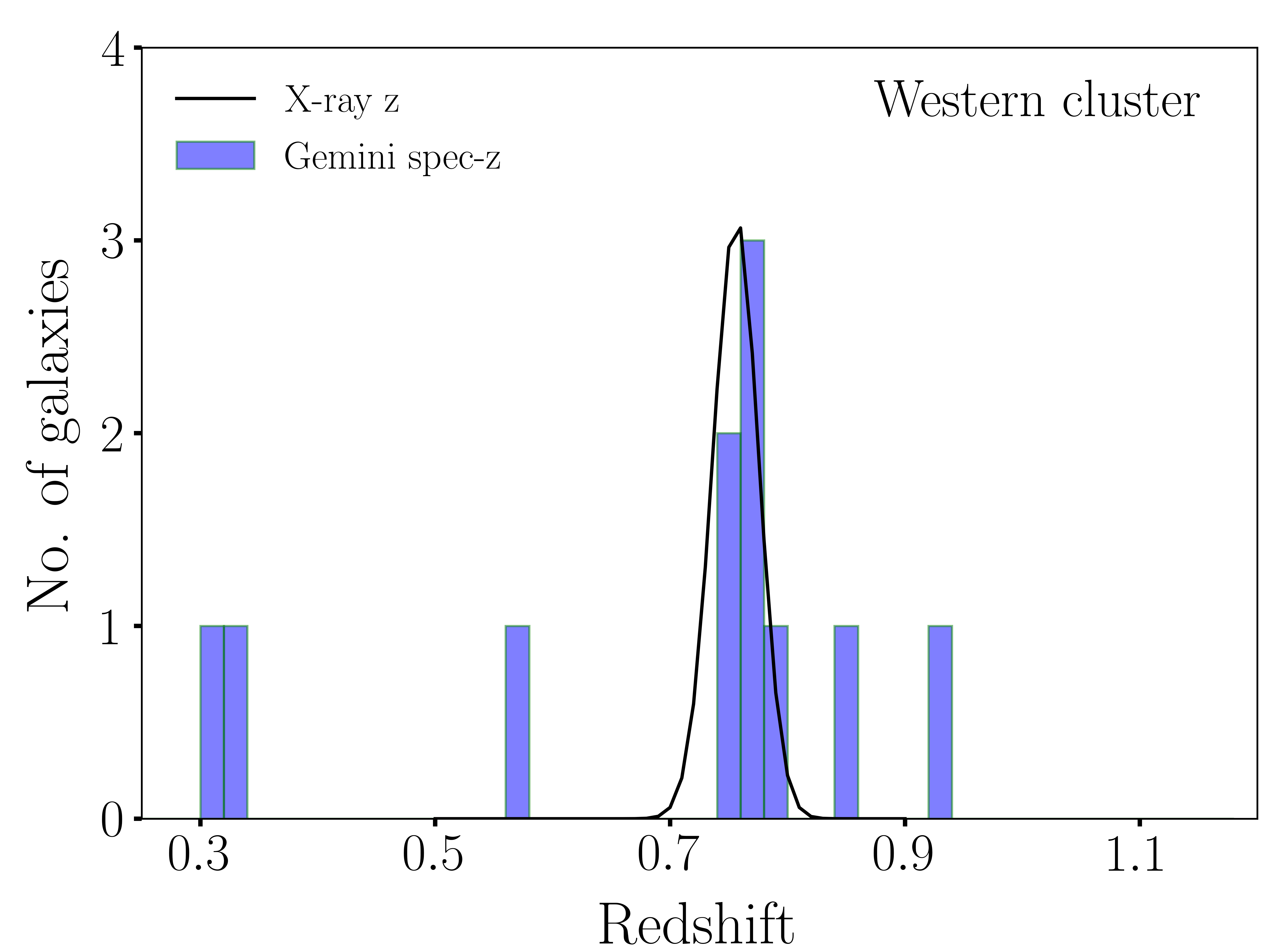}
               \caption{Spectroscopic redshift information of the two clusters. \textit{Left:} Spatial distribution of the Gemini-studied galaxies with spectroscopic $z$. The colorbar represents the $z$ value of each galaxy. The light green and purple circles represent the $1^{\prime}$ area around the Eastern and Western clusters (X-ray peaks) respectively and display the regions within which galaxies are assumed to belong to the respective cluster. Galaxies with bright yellow or deep blue colors do not belong to the cluster system. \textit{Middle}: Spectroscopic galaxy $z$ histogram for the Eastern cluster. The black line represents the X-ray $z$ normalized likelihood. \textit{Right}: Same as in middle panel, but for the Western cluster. }
        \label{redshifts}
\end{figure*}

To determine accurate redshifts for the two clusters, we implement the spectroscopic galaxy information obtained with Gemini GMOS. For each cluster, we consider only galaxies within $1^{\prime}$ from the cluster's X-ray center to avoid strong contamination from the respective neighboring cluster. The spatial distributions of the galaxies and the galaxy $z$ histograms of the two systems are displayed in Fig. \ref{redshifts}.

For the Eastern cluster we find a clear peak in the galaxy distribution around $z\approx 0.75-0.8$. Five more galaxies are projected on the same line of sight, but are not cluster members, therefore we exclude them. There is an additional galaxy at $z=0.734$, $\approx 3500-9000$ km s$^{-1}$ away from any other remaining galaxy of the cluster. We exclude this object since it is highly unlikely that it is a part of the Eastern cluster. The assigned cluster $z$ is the weighted mean\footnote{The weights are calculated as $1/\sigma_{z,i}$, where $\sigma_{z,i}$ is the $z$ measurement uncertainty per galaxy.} of the remaining 10 galaxy members, which is $z_{\text{east}}=0.771$. This is consistent with the values reported on the previous studies of the entire merging system (see Sect. \ref{cluster-description}). Finally, the cluster velocity dispersion $\sigma_{\text{vel}}$ describes the width (standard deviation) of the galaxy velocity distribution in the cluster's rest frame and can be used as a cluster mass proxy. For the Eastern cluster galaxies we measure $\sigma_{\text{vel}}=1042\pm 290$ km s$^{-1}$.

For the Western cluster, the $z$ distribution is slightly more ambiguous and with fewer galaxy members. Firstly, we exclude the four galaxies that do not lie within $0.7<z<0.84$ since it is obvious that do not belong to the Western cluster. A peak distribution of six galaxies remains with a weighted mean of $z=0.768$. The velocity dispersion of the cluster is $\sigma_{\text{vel}}=743\pm 335$ km s$^{-1}$.

%One additional galaxy lies at $z=0.841$. The peculiar velocity difference between the this galaxy and the mean $z$ of the rest is $\Delta u_{\text{pec}}=c\times \dfrac{\Delta z}{1+0.768}\approx 12,000$ km s$^{-1}$, which indicates that the $z=0.841$ galaxy is not a member of the Western cluster. Thus, its exclusion is justified.

One cannot easily determine if the $\Delta z=0.003$ difference between the two clusters implies a different cosmic distance or if it arises due to their relative peculiar motion. Assuming both clusters lie at the same distance ($z=0.769$), their relative peculiar velocity along the line of sight is $\approx 520$ km s$^{-1}$, which is a typical peculiar velocity value. Given the orientation of the double cluster system and its pre-merging state, a significant relative velocity component might also exist in the orientation of the plane, vertical to the line of sight. This is discussed further in Sect. \ref{sect-shock}. On the other hand, assuming no peculiar velocities, the comoving distance between the two clusters would be $\approx 8.5$ Mpc (considering also their angular distance). This distance would imply no physical interaction takes place between the two systems and the apparent gas bridge is the result of projection effects . However, this scenario is strongly disfavored from the analysis presented in Sect. \ref{bridge-sect}. 

As a cross-check, we also constrain the X-ray redshifts ($z_{\text{X-ray}}$) of the two clusters. For each cluster, we simultaneously fitted the X-ray spectra obtained from their $<0.5\ R_{500}$ and $0.5-1\ R_{500}$\footnote{This was done after we estimated the $R_{500}$ of each cluster as described in Sect. \ref{r500-sect}.} regions with different \texttt{apec} models but linked $z_{\text{X-ray}}$. We find $z_{\text{X-ray}}=0.771^{+0.016}_{-0.012}$ and $z_{\text{X-ray}}=0.756^{+0.019}_{-0.019}$ for the Eastern and Western cluster respectively, fully consistent with the optical spectroscopic $z$. The normalized likelihoods of the $z_{\text{X-ray}}$ values are also displayed in the middle and right panels of Fig. \ref{redshifts} as black lines.

\subsection{Radius determination}\label{r500-sect}
Next, we need to estimate the $R_{500}$ radius of both clusters. To do so, we implement an iterative process of measuring X-ray cluster properties and utilizing the cluster scaling relations presented in \citet[][B19 hereafter]{bulbul19}. The latter used a high$-z$ subsample of clusters detected in the 2500 deg$^2$ SPT-SZ survey, and as such, they studied a similar cluster population with the systems that comprise SPT-CLJ2228-5828. To estimate the $R_{500}$ of the two clusters, we use the $Y_{\text{X,CE}}-M_{\text{500}}$ relation of B19, where $Y_{\text{X,CE}}=M_{\text{gas}}\times T_{\text{X,CE}}$. $Y_{\text{X,CE}}$ is the X-ray equivalent of the Compton$-y$ parameter and a proxy for the total gas pressure, while $T_{\text{X,CE}}$ is the core-excised \T\ measured within $(0.15-1)\ R_{500}$. This relation exhibits the lowest $M_{\text{500}}$ scatter among the ones studied in B19 and thus offers the best precision for constraining $R_{500}$ (since $R_{500}\propto M_{500}^{1/3}$). To measure \Mgas, \T, and $Y_{\text{X,CE}}$, we consider only the half-circle region of the clusters opposite to the merging direction and the neighboring cluster.\footnote{To separate the two equal-area cluster sides, we consider two lines that cross each cluster's center and are perpendicular to the imaginary axis that connects the two cluster centers.} These cluster sides seem to be relatively relaxed and not significantly affected by the ongoing merging. Firstly, we assume $R_{500}=1^{\prime}$, compute $Y_{\text{X,CE}}$, and estimate a new $R_{500}$. We repeat this iteration until the $R_{500}$ estimation converges to $<5\% $ from the last iteration. For the Eastern galaxy cluster we find $R_{500}=(1.52\pm 0.05)^{\prime}$, or ($675\pm 22$ kpc), and for the Western cluster we find $R_{500}=(1.60\pm 0.05)^{\prime}$, or ($711\pm 22$ kpc). The two merging clusters have very similar radii, and therefore, masses. Their $R_{500}$ radii overlap but they do not encompass each other's cluster center. According to \citet{reiprich13}, $R_{200}\approx 1.538\ R_{500}$, and thus, $R_{200}=(2.34\pm 0.08)^{\prime}$ ($1040\pm 32$ kpc) and $R_{200}=(2.46\pm 0.08)^{\prime}$ ($1094\pm 32$ kpc) for the Eastern and Western cluster respectively. For both clusters, the $R_{200}$ radii marginally encompass the center of the other cluster. 

\subsection{X-ray profiles and galaxy cluster properties}

Knowing the $z$ and $R_{500}$ of each cluster, we can determine a wide range of their X-ray properties and derive their radial profiles. All quantities are measured in the non-merging half-circle region of each cluster, i.e., the east half of the Eastern cluster and the west half of the Western cluster. We derive the SB$(r)$, $T_{\text{X}}(r)$, pressure $P(r)$, and entropy $K(r)$ profiles. We also determine $f_{\text{X}}$, the core-included (CI) and core-excised (CE) \Lx\, \T, and \Yx, \Mgas, the average metal abundance $Z$, the X-ray concentration $c_{\text{X}}$, the characteristic pressure $P_{500}$ and entropy $K_{500}$, and the total mass $M_{500, \text{X-ray}}$ as determined by X-rays. All the cluster properties are shown in Table \ref{X-ray-table}.

%can be then estimated by the $Y_{\text{X,CE}}-M_{\text{500}}$ of \citet{bulbul19} as explained earlier. Finally, we can calculate the 

\subsubsection{Surface brightness profiles}\label{sect-SB-profile}
\begin{figure*}[hbtp]
               \includegraphics[width=0.49\textwidth]{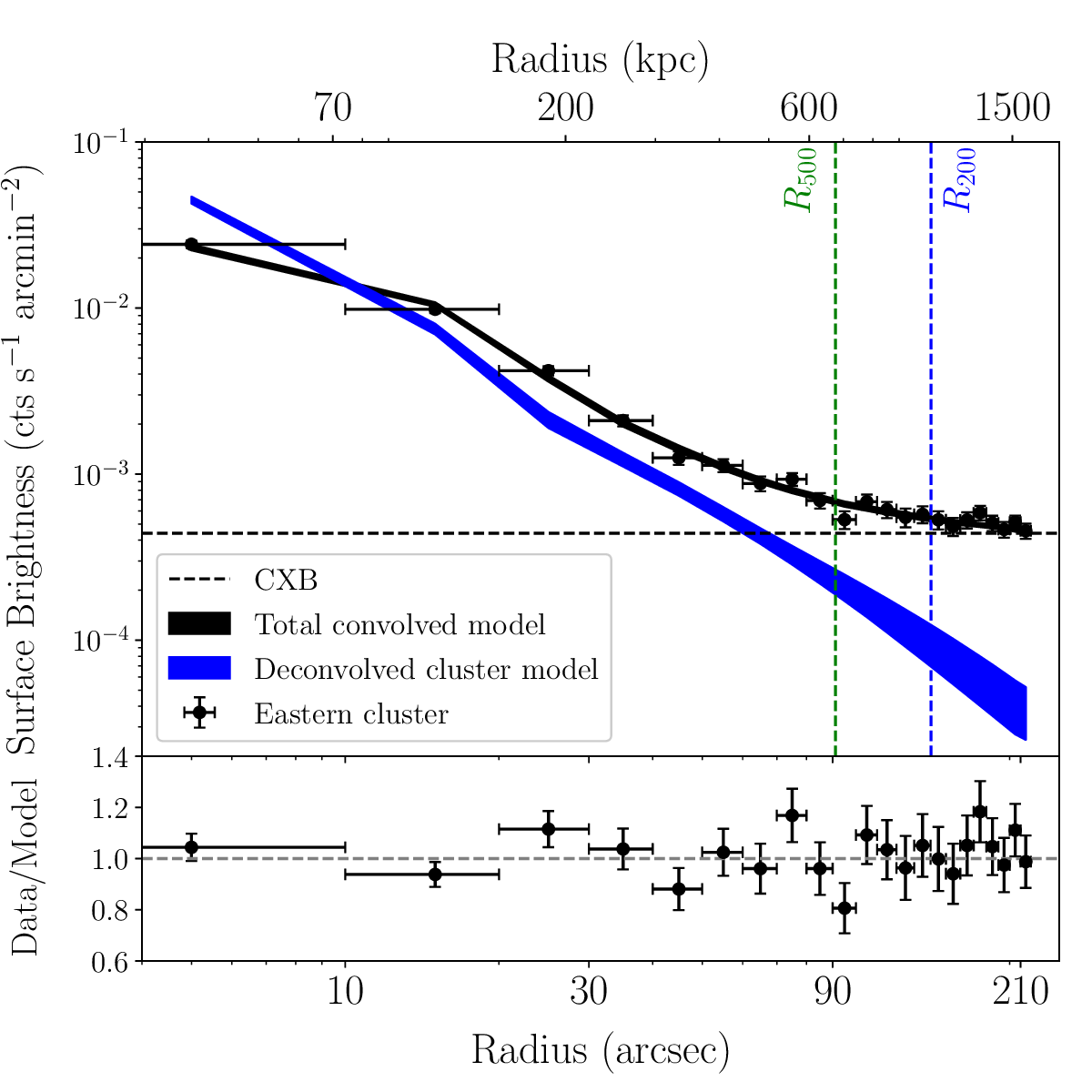}
               \includegraphics[width=0.49\textwidth]{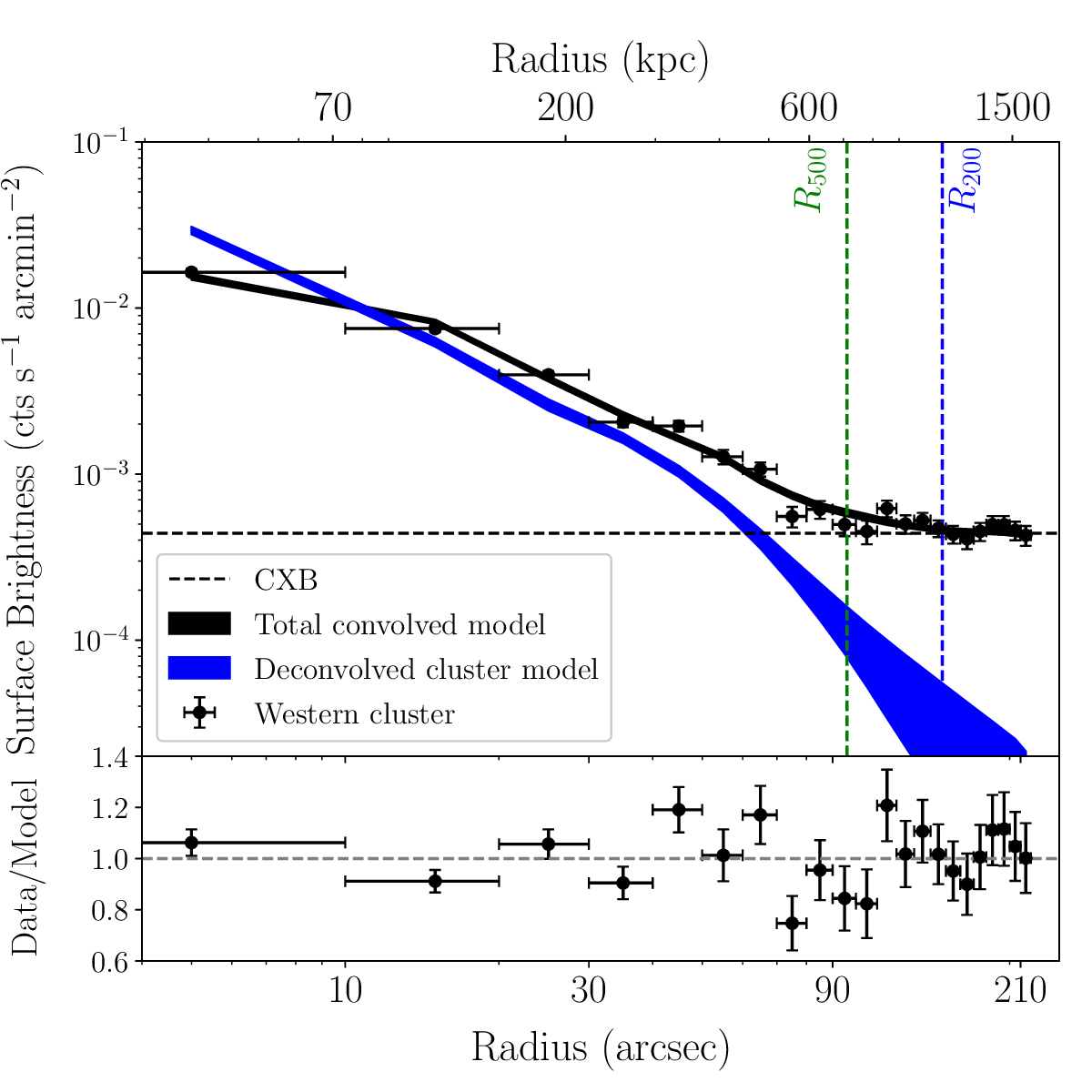}
               \caption{Surface brightness profiles (PIB-subtracted) for the Eastern and Western clusters (left and right panel respectively) in the $0.5-2$ keV band. The total (cluster$+$CXB), PSF-convolved profiles are displayed in black while the cluster-only, PSF-deconvolved profiles are shown in the filled blue band which represents the 68.3\% errorband. The $R_{500}$ and $R_{200}$ values are displayed with the dashed green and blue vertical lines respectively. The CXB level is displayed with the black horizontal line.}
        \label{SB-profiles}
\end{figure*}

The SB profiles in the $0.5-2$ keV band\footnote{All X-ray luminosities and fluxes are reported for the $0.5-2$ keV unless stated otherwise.} were extracted up to $3.5^{\prime}$, which corresponds to $\approx (2.2-2.3)\ R_{500}\approx (1.4-1.5)\ R_{200}$, or $\approx R_{100}$ for both clusters \citep{reiprich13}. The total (cluster+CXB), PSF-convolved profiles and the deconvolved cluster-only profiles are shown in Fig. \ref{SB-profiles} for both clusters. Overall, the two systems show similar SB profiles. The Eastern cluster shows a slightly more peaked central emission than the Western cluster, with its emission becoming weaker than the CXB at $\approx 0.65\ R_{500}$. Nevertheless, some excess emission is significantly detected even at $>R_{200}$. The Western cluster does not show any emission beyond $R_{200}$ and it is not as centrally peaked. However, it shows a higher SB than the Eastern cluster in the $(0.2-0.6)\ R_{500}$ range while its emission also becomes lower than the CXB at $\approx 0.65\ R_{500}$.

From the deconvolved SB profiles we determine the $0.5-2$ keV CI luminosity within $R_{500}$ to be $L_{\text{CI, east}}=(1.11\pm 0.05)\times 10^{44}$ erg s$^{-1}$ and $L_{\text{CI, west}}=(1.07\pm 0.05)\times 10^{44}$ erg s$^{-1}$ for the Eastern and Western cluster respectively. Both values correspond to $f_{\text{X}}\approx 4.5\times 10^{-14}$ erg s$^{-1}$ cm$^{-2}$. Similarly, when only the $(0.15-1)\ R_{500}$ area is considered, we measure $L_{\text{CE, east}}=(0.67\pm 0.05)\times 10^{44}$ erg s$^{-1}$ and $L_{\text{CE, west}}=(0.71\pm 0.05)\times 10^{44}$ erg s$^{-1}$ for the Eastern and Western cluster respectively. The two clusters have very similar X-ray luminosities.

Integrating the deconvolved SB profiles out to $0.1\ R_{500}$ we also measure the X-ray concentration $c_{\text{X}}=\dfrac{f_{\text{X}(<0.1\ R_{500})}}{f_{\text{X}(<R_{500})}}$ which is defined as the ratio between the emission coming from the cluster's center $0.1\ R_{500}$ over the total emission within $R_{500}$. For the two clusters we find $c_{\text{X,east}}=0.38$ and $c_{\text{X,west}}=0.31$. According to \citet{lovisari17}, the $c_{\text{X}}$ values indicate two relaxed clusters which is surprising given their actual merging state as shown in Sect. \ref{bridge-sect}. This suggests that, in cluster pairs, the opposite sides than the ones interacting probably remain unaffected, although our sample is too small to derive robust conclusions.

\subsubsection{Gas density profiles}\label{sect-ne-profile}
We extract the two $n_{\text{e}}$ profiles as described in Sect. \ref{imaging} and fit them using a double-$\beta$ model as in \citet{shitan}:
\begin{equation}
    n_{\text{e}}(r)=\left(n_{0,1}^2\left(1 + \frac{r^2}{r_{c,1}^2}\right)^{-3\beta_1} + n_{0,2}^2\left(1 + \frac{r^2}{r_{c,2}^2}\right)^{-3\beta_2} \right)^{\frac{1}{2}}
    \label{nh_eq}
\end{equation}
where $n_{0}$, $r_c$, and $\beta$, are respectively the central gas density, core scale, and slope of the two profile components. Although the double-$\beta$ model in Eq. \ref{nh_eq} is a simplified version of the $n_{\text{e}}$ profiles presented in \citet{vikhlinin06b}, it fits the $n_e$ sufficiently well.

The $n_{\text{e}}$ profiles of the two clusters are displayed in the top left panel of Fig. \ref{rest-profiles}. Their behavior compared to each other naturally resembles the SB profiles. Both clusters show a moderate central gas density of $n_{0,1}\approx (1-1.25)\times 10^{-2}$ cm$^{-3}$ with the Eastern cluster exhibiting a slightly denser core, while being marginally less dense at $(0.2-0.65)\  R_{500}$ than the Western cluster. The central $n_e$ values further suggest that the two clusters are most likely relaxed according to the relaxed-disturbed classification of \citet{lovisari17} based on several X-ray cluster properties. The scale radius of the core component is $r_{c,1}\approx 0.145\ R_{500}$ for both systems with the scale of the second component being $r_{c,2}\approx 0.58\ R_{500}$. Despite this similarity, the second components of the $n_{\text{e}}$ profiles generally differ between the two clusters, with the Western cluster showing a $72\%$ higher $n_{0,2}=(1.47\pm 0.16)\times 10^{-4}$ cm$^{-3}$ compared to the Eastern cluster, as well as a $48\%$ higher $\beta_2=0.958\pm 0.112$. The onset of the dominance of the second $n_{\text{e}}$ components is obvious for both clusters as the slopes change at $\approx (20-30)^{\prime \prime}$ . This further indicates the presence of relatively relaxed cluster cores. Furthermore, the density drops to $n_{\text{e}}\lesssim 2\times 10^{-4}$ cm$^{-3}$ at $>R_{500}$  and $n_{\text{e}}\lesssim 10^{-4}$ cm$^{-3}$ at the very outskirts ($>R_{200}$), with both clusters showing typical gas distributions. Integrating the $n_{\text{e}}$ profiles, we measure the gas masses within $R_{500}$ to be $M_{\text{gas, east}}=(1.76\pm 0.13)\times 10^{13}\ $M$_{\odot}$ and $M_{\text{gas, west}}=(2.06\pm 0.15)\times 10^{13}\ $M$_{\odot}$ for the Eastern and Western cluster respectively. This further confirms the similarity of the two clusters and their average mass.

\subsubsection{Temperature profiles and metal abundance}\label{sect-T-profile}
Firstly, we determine the overall \T\ values for both clusters. For the Eastern cluster, we find \T$_{,CI}=3.64^{+0.43}_{-0.35}$ keV and \T$_{,CE}=3.22^{+0.59}_{-0.50}$ keV for the core-included and core-excised regions respectively. Similarly, for the Western cluster we find \T$_{,CI}=3.84^{+0.36}_{-0.35}$ keV and \T$_{,CE}=3.41^{+0.54}_{-0.45}$ keV. The two clusters show very similar \T, with the Western cluster being slightly hotter, which is consistent with the marginally higher values of $L_{\text{CE}}$ and \Mgas\ it also shows compared to the Eastern cluster.

The universal metal abundance $Z$ could be only loosely constrained with $Z_{\text{east}}=0.84^{+0.42}_{-0.37}\ Z_{\odot}$ and $Z_{\text{west}}=0.76^{+0.36}_{-0.31}\ Z_{\odot}$. Although the best-fit values are relatively high, the large uncertainties prevent us from drawing firm conclusions. For all following \T\ constraints, we let $Z$ free to vary within the $1\sigma$ limits of the $<R_{500}$ constraints.

Due to the small apparent size of the two clusters, extracting and fitting \T\ profiles is challenging. To have a sufficient number of counts per bin, avoid strong photon mixing between bins due to the limited PSF, and have enough bins to constrain the \T\ profile, we define three bins for each cluster: $0-0.3\ R_{500}$, $0.3-0.6\ R_{500}$, and $0.6-1\ R_{500}$. For each bin, we measure the 2D projected \T. Once again, the two cluster systems show very similar behavior. A gradual \T\ decrease is observed with the central, middle, and outer bins showing $\approx 3.8-3.9$ keV,  $\approx 3-3.3$ keV and $\approx 2-2.1$ keV respectively, for both clusters. 

\begin{figure*}[h]
\centering
               \includegraphics[width=0.45\textwidth]{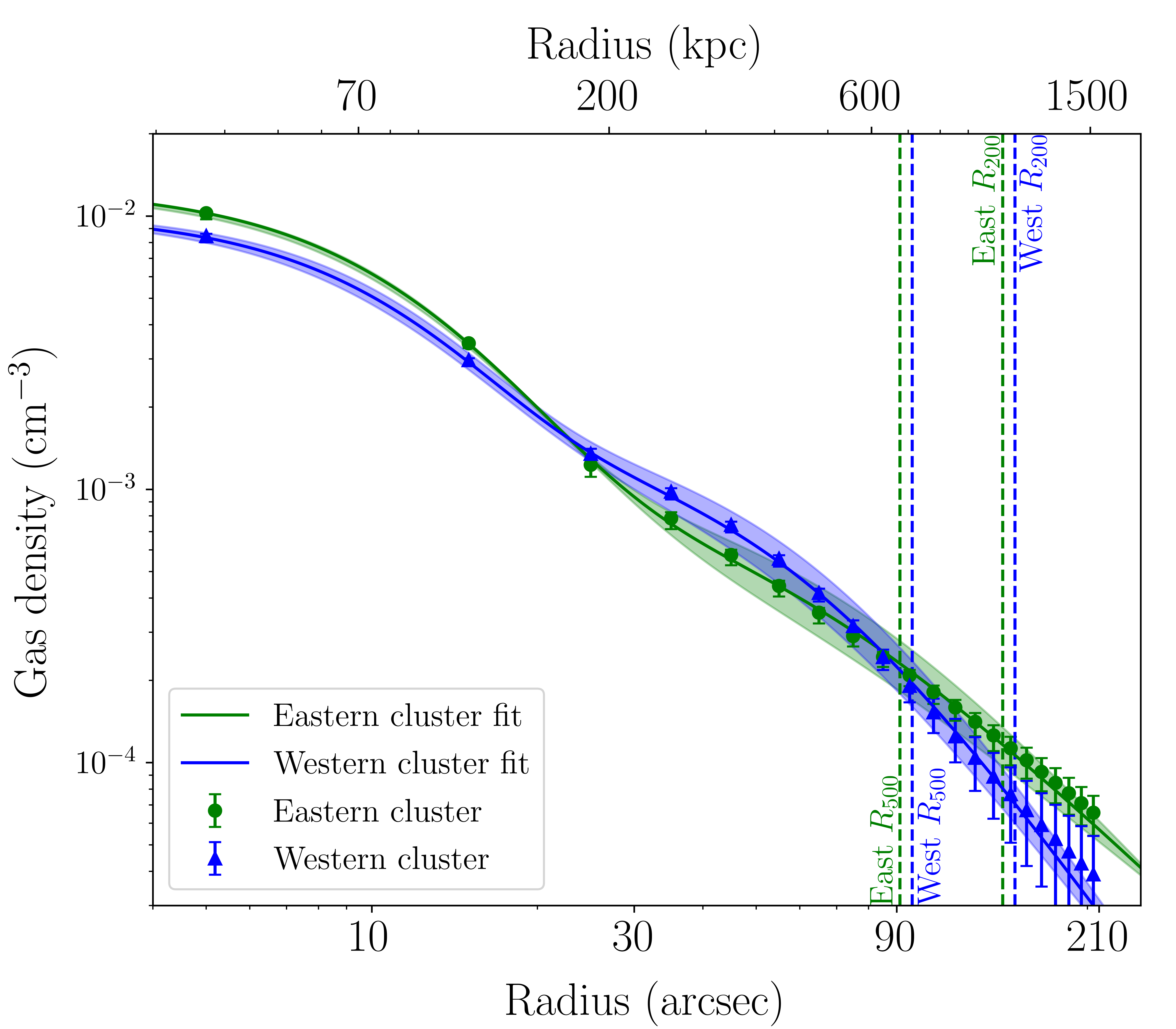}
               \includegraphics[width=0.45\textwidth]{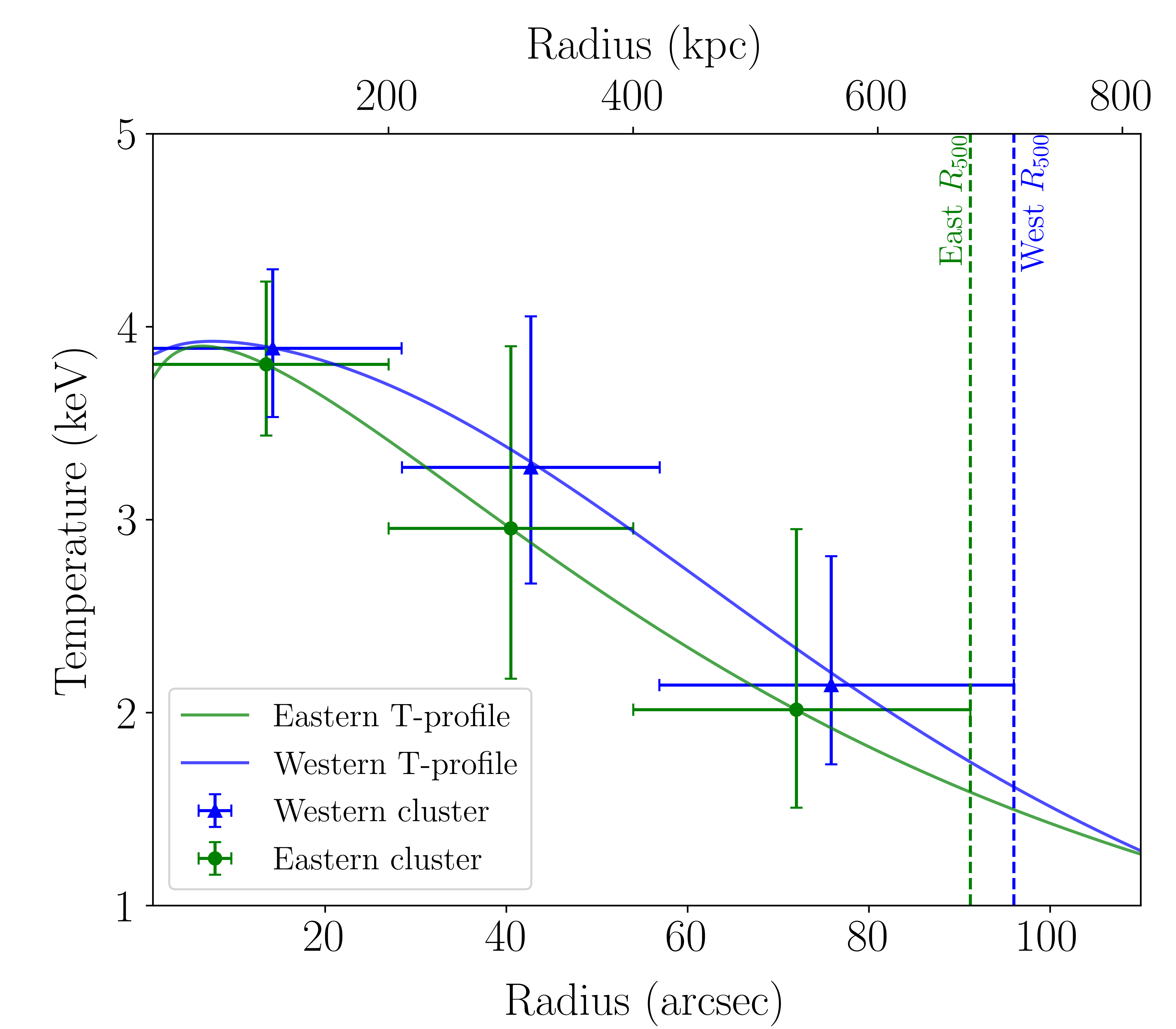}
               \includegraphics[width=0.45\textwidth]{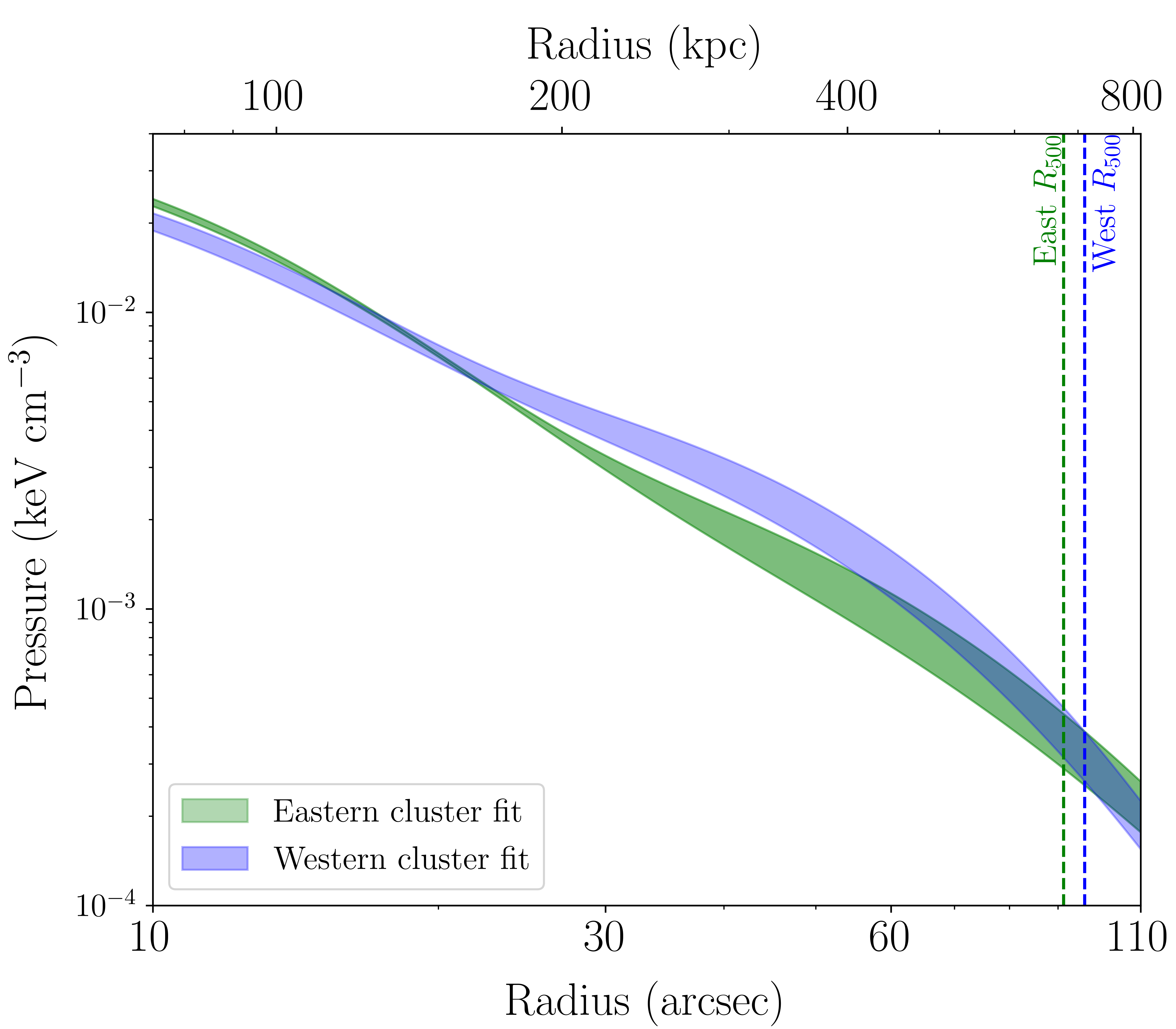}
               \includegraphics[width=0.45\textwidth]{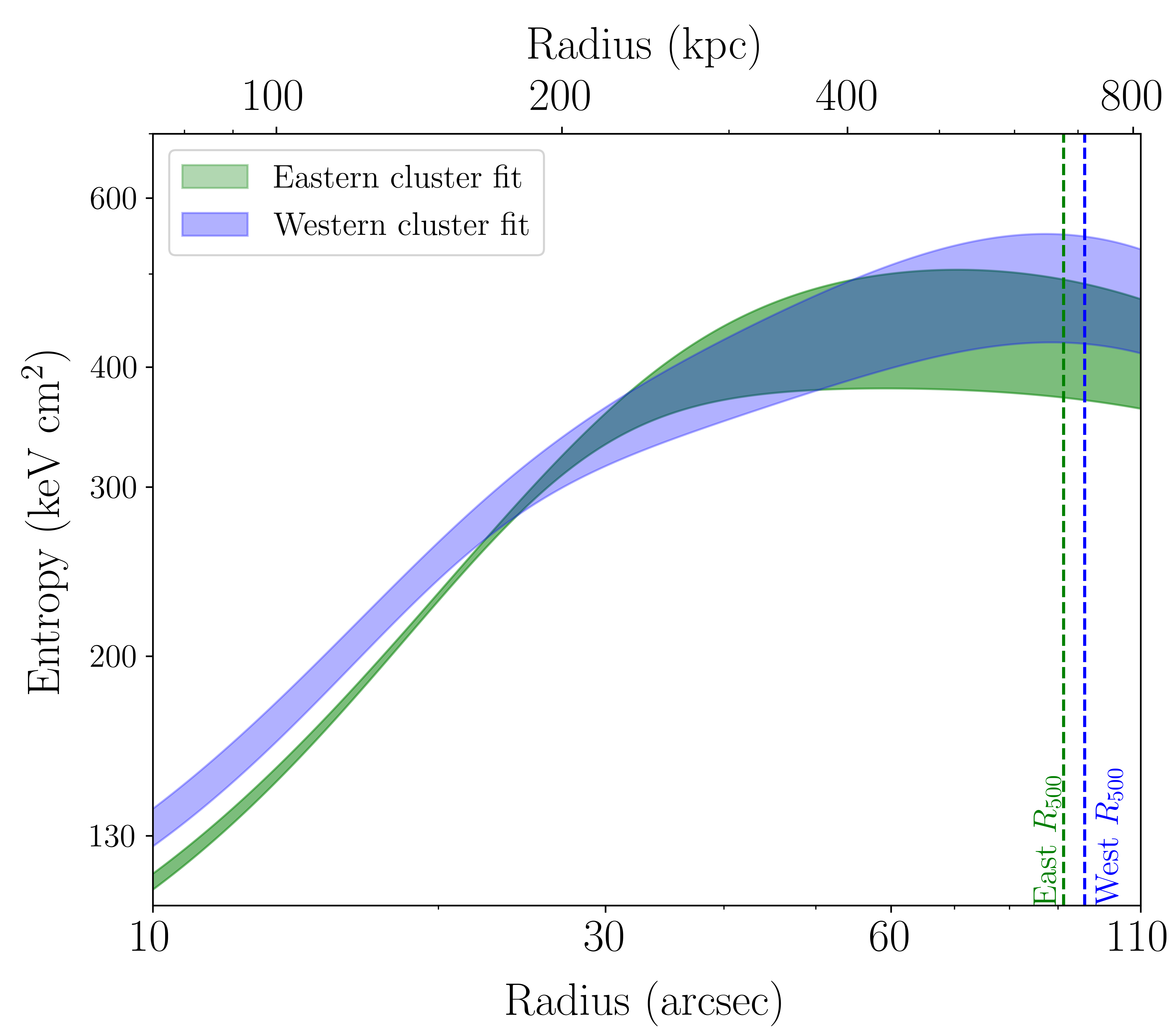}
               \caption{X-ray radial profiles for the Eastern and Western clusters (green and blue colors respectively). The $R_{500}$ values are displayed with the dashed vertical lines. \textit{Upper left:} Gas (electron number) density $n_e$ profiles with their best-fit models. The $R_{200}$ values are also displayed in this panel. \textit{Upper right:} Temperature profiles (projected, 2D) with their best-fit deprojected, 3D models. Due to the limited data, no model uncertainties could be derived. \textit{Bottom left:} Inferred pressure profile best-fit models which are computed as $P=n_e\times T_{\text{X,3D}}$. \textit{Bottom right:} Inferred entropy profile best-fit models which are computed as $K=T_{\text{X,3D}}\times n_e^{-\frac{2}{3}}$.}
        \label{rest-profiles}
\end{figure*}

To determine the 3D deprojected $T_{\text{X,3D}}(r)$ profiles we project the latter onto the line of sight and fit the measured \T. To do so, we adopt the $T_{\text{X,3D}}(r)$ profile described in \citet{vikhlinin06b} and the projection weighting scheme introduced by \citet{vikhlinin06a}. The $T_{\text{X,3D}}(r)$ parameters are then determined by considering the resulted projected $T_{\text{X,2D}}(r)$ profile that describes the data best. This methodology has been implemented by several past studies \citep[see, e.g.,][for more details]{bartalucci, gerrit2, lovisari20}. The \T$_{,3\text{D}}(r)$ profile is given by:
\begin{equation}
    T_{\text{X,3D}}(r) = T_{\text{X,CE}}^* T_{\text{max}} \frac{\left(r/r_{\text{cool}}\right)^{a_{\text{cool}}} + T_0/T_{\text{max}}}{\left(r/r_{\text{cool}}\right)^{a_{\text{cool}}} + 1} \frac{\left(r/r_t\right)^{-a}}{\left(1 + \left(r/r_t\right)^b\right)^{c/b}}
    \label{T-profile-eq}
\end{equation}
where $T_{\text{X,CE}}^*$ is the \T\ measured within $(0.15-0.75)\ R_{500}$.\footnote{Throughout the paper we take $T_{\text{X,CE}}$ to be measured within $(0.15-1)\ R_{500}$ for consistency with B19. However, in the $T_{\text{X,3D}}(r)$ model, it is considered to be measured within $(0.15-0.75)\ R_{500}$ for consistency with other studies that used this model. This is the reason we depict it with a "*" symbol, $T_{\text{X,CE}}^*$.} Moreover, $T_0$, $r_{\text{cool}}$, and $a_{\text{cool}}$ drive the behavior of the $T_{\text{X,3D}}(r)$ profile close to the core ($\lesssim [0.05-0.15]\ R_{500}$), while $T_{\text{max}}$, $r_t$, $b$, and $c$ describe the profile further out until the outskirts. Since we only have three measured \T\ bins, this eight-parameter model will clearly overfit the data. To this end, we fix $(r_{\text{cool}}, a,
a_{\text{cool}}, r_t, b)=(0.039\ R_{500}, 0.036, 0.898, 1.095\ R_{500}, 2.36)$ to the best-fit values obtained by \citet{chen23} for the EXCPReS (Evolution of X-ray galaxy Cluster Properties in a Representative Sample) catalog, which contains $0.4<z<0.6$ clusters with similar properties as the ones we focus in this work. Also, \citet{chen23} constrained $T_0=0.509\pm 0.452$ but we fix it to $T_0=0.9$ which is more suitable for our two clusters.\footnote{The two values are consistent within $1\sigma$ while our fitted profiles are insensitive to $T_0$ as we could not measure \T\ so close to the core. Thus, the exact $T_0$ value is only important when we infer $P$ and $K$ profiles later on. Moreover, since our clusters are at higher-$z$ than EXCPReS, any cool \T\ gradients toward the cluster core are likely expected to be weaker \citep[e.g.,][]{santos10} for our clusters, implying a higher $T_0$. Finally, from the CI and CE \T\ values, it seems unlikely that $T_{\text{X,3D}}$ drops significantly below $T_{\text{X,CE,0.75}}$ toward the core. All the above justify our choice of $T_0=0.9$.}  Thus, we only fit the $T_{\text{max}}$ and $c$ parameters which have the most impact on our $T_{\text{X,3D}}$ profiles. The projected best-fit \T\ profiles, alongside the \T\ measurements, are displayed in the top right panel of Fig. \ref{rest-profiles}. The statistical uncertainties of the \T$_{,3\text{D}}(r)$ profiles cannot be robustly constrained due to the limited data and the wide range of model parameters, many of which we keep fixed, which underestimates the uncertainties. Therefore, we only focus on the best-fit \T\ profile functions.
The two clusters return very similar results near their center ($<0.3\ R_{500}$) and toward the outskirts ($>0.65\ R_{500}$). In the in-between cluster regions the Western cluster seems to be $\approx 11\%$ hotter, although this is well within the statistical uncertainties. Even if very limited information on the \T-profiles is available, the generally used profile in Eq. \ref{T-profile-eq} gives a very good description of the data even with most parameters fixed to literature values. As such, to derive the pressure and entropy profiles in the next sections, we use the best-fit functions of the \T\ profiles, assuming no uncertainties.

\subsubsection{Pressure profiles, X-ray Compton equivalent $Y_{\text{X}}$, and total mass $M_{500, Y_{\text{X}}}$}

As explained in Sect. \ref{r500-sect}, the X-ray equivalent for the Compton parameter is defined as $Y_{\text{X}}=M_{\text{gas}}\times T_{\text{X}}$ and is a proxy of the total gas pressure of the cluster. The CI and CE $Y_{\text{X}}$ values are computed by the use of $T_{\text{X,CI}}$ and $T_{\text{X,CE}}$ respectively. For the Eastern cluster we find $Y_{\text{X,CI}}=6.40^{+0.92}_{-0.80}\times 10^{13}\ M_{\odot}$ keV and $Y_{\text{X, CE}}=5.67^{+1.12}_{-1.01}\times 10^{13}\ M_{\odot}$ keV while for the Western cluster we find $Y_{\text{X,CI}}=7.91^{+0.96}_{-0.93}\times 10^{13}\ M_{\odot}$ keV and $Y_{\text{X, CE}}=7.01^{+1.23}_{-1.07}\times 10^{13}\ M_{\odot}$ keV. The total mass within $R_{500}$ as inferred from the $Y_{\text{X,CE}}-M_{500}$ relation of B19 (after correcting for the different cosmology) is $M_{500,Y_{\text{X}}}=(2.10\pm 0.20)\times 10^{14}\ M_{\odot}$ and $M_{500,Y_{\text{X}}}=(2.41\pm 0.22)\times 10^{14}\ M_{\odot}$ for the Eastern and Western cluster respectively.\footnote{We do not measure the hydrostatic mass of the clusters since we do not have realistic uncertainties for the \T\ profile, which would result in strongly underestimated uncertainties of the hydrostatic mass.}

The cluster gas pressure is the least affected thermodynamic property by the thermodynamic evolution of clusters. The gas pressure profile, assuming an ideal and monoatomic gas, is given by $P(r)=n_{\text{e}}(r)T_{\text{X,3D}(r)}$. Since we have derived both the $n_{\text{e}}(r)$ and $T_{\text{X,3D}(r)}$ profiles, we take their product to be the $P(r)$ profiles. The latter are shown in the bottom left panel of Fig. \ref{rest-profiles}. Due to the limited \T\ information, we cannot infer the $P(r)$ profile much further than $R_{500}$. Nevertheless, we see that both $P(r)$ profiles show a typical behavior for clusters of such \T\ and $z$ \citep[e.g.,][]{ghirardini17,He21}. According to the results of \citet{arnaud10}, the $P(r)$ profiles indicate averagely relaxed clusters without strong cool core presence (but possible the presence of weak cool cores). Moreover, under the assumption of self-similarity, the characteristic pressure $P_{500}$ of a cluster is given by \citep{voit05,nagai}:
\begin{equation}
    P_{500}=1.89\times 10^{-3}\ E(z)^{\frac{8}{3}}\times \left(\dfrac{M_{500}}{3\times 10^{14}\ M_{\odot}}\right)^{\frac{2}{3}}\ h^{8/3}\ \text{keV}/\text{cm}^{3} 
\end{equation}
where $E(z)=\sqrt{\Omega_{\text{m}}(1+z)^3+\Omega_{\Lambda}}$ and $h=H_0/(70$\ km s$^{-1}$ Mpc$^{-1}$). For the Eastern and Western cluster we find $P_{500}=(4.92\pm 0.33)\times 10^{-3}$ keV cm$^{-3}$ and $P_{500}=(5.42\pm 0.32)\times 10^{-3}$ keV cm$^{-3}$ respectively, which approximately corresponds to the $P(r)$ value at $r\approx 0.3\ R_{500}$, consistent with \citet{ghirardini17} within the scatter of clusters at similar $z$. Finally, we do not have sufficient data to probe the previously detected flattening of the profile at $<0.1\ R_{500}$ \citep[e.g.,][]{mcdonald13}.

\subsubsection{Entropy profiles}
The entropy $K$ of the ICM is an essential tool for studying the thermodynamic state and history of the cluster gas. It always increases when gas heating occurs \citep[e.g.,][]{voit05} and can help us understand the relaxation state of the cluster core and outskirts, as well as their gas clumpiness \citep[e.g.,][]{bulbul16}. The entropy profile is given by $K(r)=T_{\text{X,3D}}(r)/ n_{\text{e}}(r)^{2/3}$, and thus, it can be directly inferred from the known $T_{\text{X,3D}}(r)$ and $n_{\text{e}}(r)$ profiles. The $K(r)$ profiles are shown in the bottom right panel of Fig. \ref{rest-profiles}. As discussed before, they are only loosely constrained due to the limited \T\ information. Nevertheless, one sees that $K(r)$ decrease toward the cluster center at a level expected for relatively average clusters in terms of relaxation state \citep[$K\approx 100-130$ keV cm$^2$ at the core, e.g., ][]{cavagnolo, ghirardini17}. Toward the outskirts, one sees that the slope of the $K(r)$ profiles flatten which indicates that the cluster outskirts are not yet in dynamical equilibrium (although no robust conclusions can be drawn due to the large uncertainties). Under the self-similar model assumption, the characteristic entropy $K_{500}$ of the clusters is \citep{voit05b}:
\begin{equation}
    K_{500}=761\times E(z)^{-\frac{2}{3}}\times \left(\dfrac{M_{500}}{3\times 10^{14}\ M_{\odot}}\right)^{\frac{2}{3}}\ h^{\frac{2}{3}}\ \text{keV}\ \text{cm}^{2}. 
\end{equation}
For the Eastern and Western cluster we find $K_{500}=451\pm 31$ keV cm$^{-3}$ and $K_{500}=498\pm 30$ keV cm$^{-3}$ respectively, which approximately corresponds to the $K(r)$ value at $r\approx (0.75-0.8)\ R_{500}$.

\subsection{Weak lensing masses}
The two-component NFW fit to the weak lensing data results in the following masses for the cluster merging system: $M_{200}=3.26^{+1.25}_{-1.24}\times10^{14} M_{\odot}$ for the Western cluster, $M_{200}=0.56^{+0.80}_{-0.49}\times10^{14} M_{\odot}$ for the Eastern cluster, and $M_{\text{tot}}=3.82^{+1.90}_{-0.97}\times10^{14} M_{\odot}$ for the total mass of the system. The latter was determined by adding the $M_{200}$ of the cluster components for each step in the MCMC. We also obtain $M_{500}=2.21^{+0.86}_{-0.85}\times10^{14} M_{\odot}$ for the Western cluster and $M_{500}=0.40^{+0.53}_{-0.34}\times10^{14} M_{\odot}$ for the Eastern cluster. Here the quoted WL mass uncertainties include shape noise only (see Sect.\thinspace\ref{se:compare_lensing_xray} regarding the impact of additional noise components).

Interestingly, the Eastern cluster is marginally detected by the weak lensing data, only at a $1.14\sigma$ level. Moreover, the Western cluster appears to be $\approx 6$ times more massive compared to the Eastern component (although only at $1.8\sigma$), in contrast with the similarity the two clusters exhibit in X-rays. Finally, the total mass of the system (sum of the two $M_{200}$ values) is consistent with the past SZ constraints on the total mass as presented in Sect. \ref{cluster-description}.

%\begin{figure}[h] 
%\centering
%\includegraphics[width=0.45\textwidth]{degen.bcgwest.bcgeast.pdf}
%\caption{\label{fig:degen.bcgcebtr} $M_{200}$ probability contours and degeneracy of the two mass components in the two-component NFW fit from the WL analysis.} 
%\end{figure}

\begin{table*}[htbp]
\centering
\caption{X-ray and optical properties of the Eastern and Western clusters. (1) Right Ascension J2000. (2) Declination J2000. (3) Optical spectroscopic redshift from Gemini galaxies. (4) X-ray spectroscopic redshift. (5) X-ray luminosity in the $0.5-2$ keV band measured within $<R_{500}$ (core-included) in $10^{44}$ erg s$^{-1}$ units. (6) Same as (5) but measured within $(0.15-1)\ R_{500}$. (7) ICM temperature measured within $<R_{500}$ (core-included) in keV units. (8) Same as (7) but measured within $(0.15-1)\ R_{500}$. (9) Metal abundance measured within $<R_{500}$ in $Z_{\odot}$ units. (10) Gas mass measured within $<R_{500}$ in ($10^{13}\ M_{\odot}$) units. (11) X-ray equivalent of the Compton$-y$ parameter measured within $<R_{500}$ in $10^{13}\ M_{\odot}\ $keV units. (12) Same as (11) but measured within $(0.15-1)\ R_{500}$. (13) Total mass estimated within $<R_{500}$ in $10^{14}\ M_{\odot}$ units using the $Y_{\text{X,CE}}-M_{500}$ relation of B19.(14) Total mass within $<R_{200}$ in $10^{14}\ M_{\odot}$ units using the $R_{200}=1.538\ R_{500}$ relation of \citet{reiprich13} and the determined $R_{500}$ value. (15) $R_{500}$ radius in kpc units. (16) $R_{200}$ radius in kpc units. (17) Characteristic pressure in $10^{-3}$ keV cm$^{-3}$ units. (18)  Characteristic entropy in keV cm$^2$ units. (19) Total mass measured within $<R_{500}$ in $10^{14}\ M_{\odot}$ units from the WL data. (20) Same as in (19) but within $<R_{200}$. (21) Velocity dispersion measured in km s$^{-1}$ units. (22) X-ray concentration parameters.}
\label{X-ray-table}
\begin{tabular}{lcc}
\hline
\hline
Cluster property & Eastern Cluster & Western Cluster \\
\hline
(1) R.A. & 337.239$^{\circ}$ & 337.171$^{\circ}$ \\
(2) DEC. & $-58.479^{\circ}$ & $-58.474^{\circ}$ \\
(3) $z_{\text{opt}}$ & 0.771 & 0.768 \\
(4) $z_{\text{X-ray}}$ & $0.771^{+0.016}_{-0.012}$ & $0.756^{+0.019}_{-0.019}$ \\
(5) $L_{\text{X,CI}}$ & $1.11\pm 0.05$ & $1.07\pm 0.05$ \\
(6) $L_{\text{X,CE}}$ & $0.67\pm 0.05$ & $0.71\pm 0.05$ \\
(7) $T_{\text{X,CI}}$ & $3.638^{+0.429}_{-0.351}$ & $3.841^{+0.355}_{-0.351}$ \\
(8) $T_{\text{X,CE}}$ & $3.219^{+0.592}_{-0.497}$ & $3.405^{+0.538}_{-0.451}$ \\
(9) $Z$ & $ 0.837^{+0.418}_{-0.372}$ & $ 0.764^{+0.359}_{-0.310}$ \\
(10) $M_{\text{gas}}$ & $1.76\pm 0.13$ & $2.06\pm 0.15$ \\
(11) $Y_{\text{X,CI}}$ & $6.40^{+0.92}_{-0.80}$ & $7.91^{+0.96}_{-0.93}$\\
(12) $Y_{\text{X,CE}}$ & $5.67^{+1.12}_{-1.01}$ & $7.01^{+1.23}_{-1.07}$ \\
(13) $M_{500, Y_{\text{X}}}$ & $2.10\pm 0.20$ & $2.41\pm 0.22$ \\
(14) $M_{200, Y_{\text{X}}}$ & $3.06\pm 0.29$ & $3.51\pm 0.32$ \\
(15) $R_{500}$  & $675\pm 22$ & $711\pm 22$  \\
(16) $R_{200}$  & $1040\pm 34$ & $1094\pm 34$  \\
(17) $P_{500}$ & $4.92\pm 0.33$ & $5.42\pm 0.32$ \\
(18) $K_{500}$ & $451\pm 31$ & $498\pm 30$ \\
(19) $M_{500, \text{WL}}$ & $0.40^{+0.53}_{-0.34}$ & $2.21^{+0.86}_{-0.85}$\\
(20) $M_{200, \text{WL}}$ & $0.56^{+0.80}_{-0.49}$ & $3.26^{+1.25}_{-1.24}$ \\
(21) $\sigma_{\text{vel}}$ & $1042\pm 290$ & $743\pm 335$ \\
(22) $c_{\text{X}}$ & 0.38 & 0.31 \\
\hline
\end{tabular}
\end{table*}

$M_{200, Y_{\text{X}}}=(3.06\pm 0.29)\times 10^{14}\ M_{\odot}$ and $M_{200, Y_{\text{X}}}=(3.51\pm 0.32)\times 10^{14}\ M_{\odot}$

\subsection{Comparison of X-ray and WL cluster masses}\label{mass-consist}
\label{se:compare_lensing_xray}

The X-ray masses are estimated from the $Y_{\text{X,CE}}-M_{500}$ relation of B19. The $M_{500}$ values in that work were inferred from the unbiased SZ S/N ratio $\zeta$ and the $\zeta-M_{500}$ scaling relation of \citet{dehaan16}. The latter calibrated the $\zeta-M_{500}$ relation using masses obtained by $Y_{\text{X}}$ values measured by \textit{Chandra} and the $Y_{\text{X}}-M_{500}$ from \citet{vikhlinin09,vikhlinin09b}, where $M_{500}$ represented the hydrostatic masses. The normalization of the latter $Y_{\text{X}}-M_{500}$ relation was adjusted in \citet[][see their Sect. 4.2 for more details]{dehaan16} so the resulted $M_{500}$ were consistent with more recent WL mass estimates from \citet{hoekstra15}. This way they accounted for any systematic offset between the WL and the original hydrostatic $M_{500}$. Given this sequence of scaling relation calibrations, and assuming no significant biases were propagated through these relations, we expect that the derived $M_{500, Y_{\text{X}}}$ values in this work are consistent with the WL mass estimates.

To compare the WL and X-ray cluster masses, we assume that $R_{500}=1.538\ R_{200}$ \citep{reiprich13}, which results in $M_{200, Y_{\text{X}}}=(3.06\pm 0.29)\times 10^{14}\ M_{\odot}$ and $M_{200, Y_{\text{X}}}=(3.51\pm 0.32)\times 10^{14}\ M_{\odot}$ for the Eastern and Western clusters respectively. The Western cluster mass estimates agree well between WL and X-ray within $<0.2\sigma$, with $M_{200, \text{WL}}=0.93_{-0.30}^{+0.50}\ M_{200, Y_{\text{X}}}$. On the other hand, the Eastern cluster is marginally detected in WL and, as a result, the two mass estimates disagree at a $2.9\sigma$ level, with $M_{200, \text{WL}}=0.18^{+0.28}_{-0.16}\ M_{200, Y_{\text{X}}}$. Even though its statistical uncertainties are large, the $M_{200, \text{WL}}$ value of the Eastern cluster is surprisingly low. Due to the similarity of the two clusters in their X-ray properties, if there was a true lack of WL signal from the Eastern cluster it would be striking and highly interesting for a follow-up study. However, we stress that the (already large) quoted WL mass uncertainty includes shape noise only. Additional statistical uncertainties are caused by large-scale structure projections and line-of-sight variations in the source redshift distribution. For individual clusters at the mass and redshift range of the target and analyses of similar setup these jointly amount to an additional statistical uncertainty of \mbox{$\Delta M_{\mathrm{200c}, \text{WL}} \simeq 0.8\times 10^{14}\ M_{\odot}$} \citep[e.g.][]{schrabback18}, which would need to be added in quadrature to the comparable shape noise uncertainties. In the special case of two clusters  located in close proximity on the sky, these noise contribution would however be correlated. This is expected to reduce the relative impact of these uncertainties on the mass ratio of the components, but makes it  difficult to quantify the impact precisely. Nevertheless, it is clear that the true significance of the discrepancy between the X-ray and WL mass estimates of the Western component is lower than what is estimated above, making it plausible that this could simply be the result of a statistical fluke.
%However, we believe that this is probably caused by an unidentified systematic uncertainty in the $M_{200, \text{WL}}$ of the Eastern cluster.

\subsection{Consistency with known scaling relations of single clusters}\label{comparison-scaling}

Merging clusters are often excluded from samples constructed to constrain average scaling relations \citep[e.g.,][]{migkas20} since the merging process is expected to alter the scaling relation behavior of such clusters. X-ray and SZ scaling relations are mostly driven by the gravitational effects on the ICM. At the same time, the ICM behavior in merging clusters is strongly affected by formed shock fronts and general gas disruption. However, limited effort has been shown to evaluate if the outer, undisturbed side of pre-merger systems follow an average scaling relation behavior. To assess this using the two clusters studied in this work, we estimate the consistency of their X-ray and WL properties as measured from their undisturbed sides with the scaling relations presented by B19. In that work the authors used only isolated, non-merging clusters to constrain scaling relations between X-ray observables and $M_{500}$. Using $M_{500, \text{WL}}$ for these comparisons,  we find that the Western cluster lies within $<1\sigma$ for all seven scaling relations studied in B19. For the Eastern cluster, such comparisons are not feasible due to the marginal detection of the cluster in the WL data and its likely underestimated $M_{500, \text{WL}}$ value.

To compare the consistency between the clusters' X-ray properties and the B19 scaling relations, we perform the following. Based on the B19 scaling relations, we infer $M_{500}$ with all available X-ray properties (\Mgas, CI and CE \Lx, \T, and \Yx) and compare the results. For example, if $M_{500}$ from \Yx\ and \Lx\ agree within the statistical scatter, then the studied cluster is considered consistent with the B19 scaling relations, i.e., it would lie close to the hypothetical $Y_{\text{X}}-L_{\text{X}}$ scaling relation based on the B19's measurements. For the Western cluster, we find that $M_{500}$ estimates from all X-ray properties agree within $1.2\sigma$ (given the $M_{500}$ scatter of each relation). The most discrepant quantities are $T_{\text{X,CE}}$ and \Mgas\ which predict a $M_{500}$ difference of $\approx 20\pm 17\%$. For the Eastern cluster, we find similar results with a general $M_{500}$ agreement within $1.1\sigma$ and the most discrepant quantities $L_{\text{X, CI}}$ and \Mgas\ predicting different $M_{500}$ by $\approx 30\pm 27\%$. 

Clearly, we do not have a sufficient number of clusters to draw robust conclusions while the $M_{500}$ scatter of the B19 scaling relations is rather large, covering most of the inferred $M_{500}$ differences. Nonetheless, the results of both clusters favor the case that the undisrupted side of clusters at the early merging stages can be used to measure X-ray properties that would correspond to isolated, undisturbed clusters. This potentially allows the use of such merging clusters as single, isolated clusters, as long as their merging side is ignored.

\subsection{Consistency with the SPT-SZ observations and cosmological implications}

\begin{figure}[h]
\centering
               \includegraphics[width=0.4\textwidth]{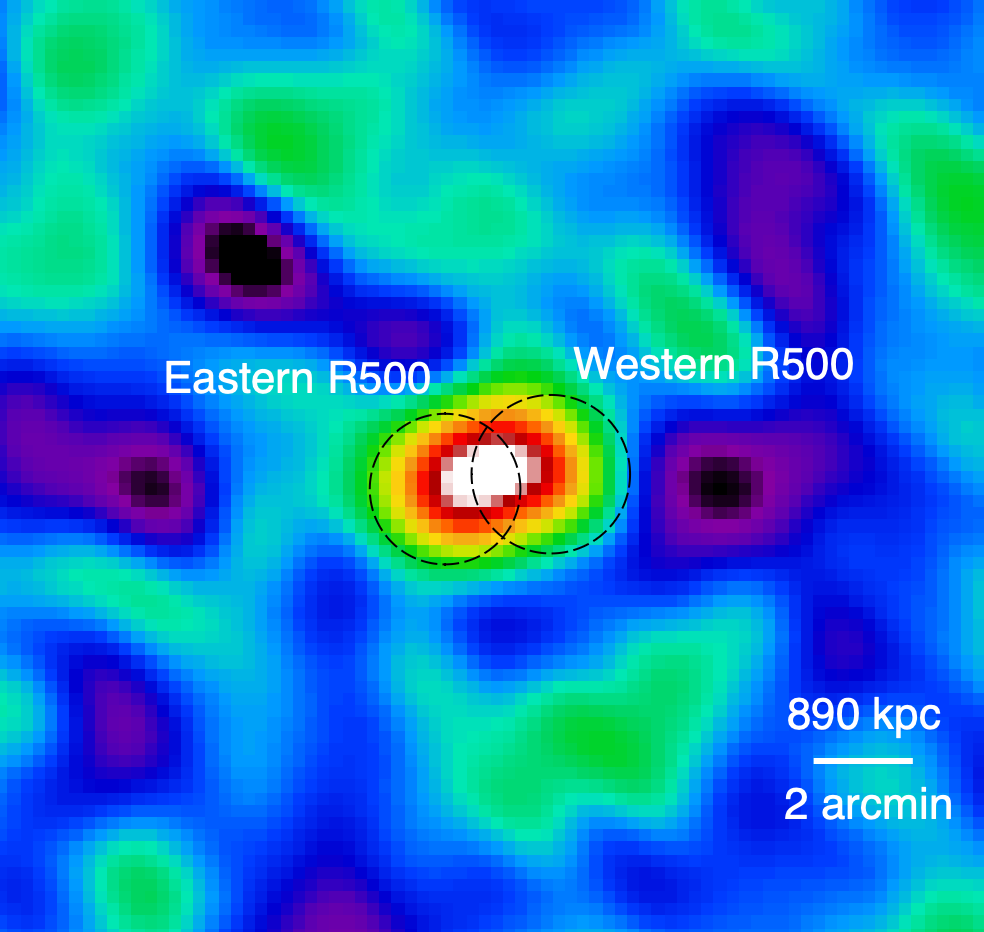}
               \includegraphics[width=0.4\textwidth]{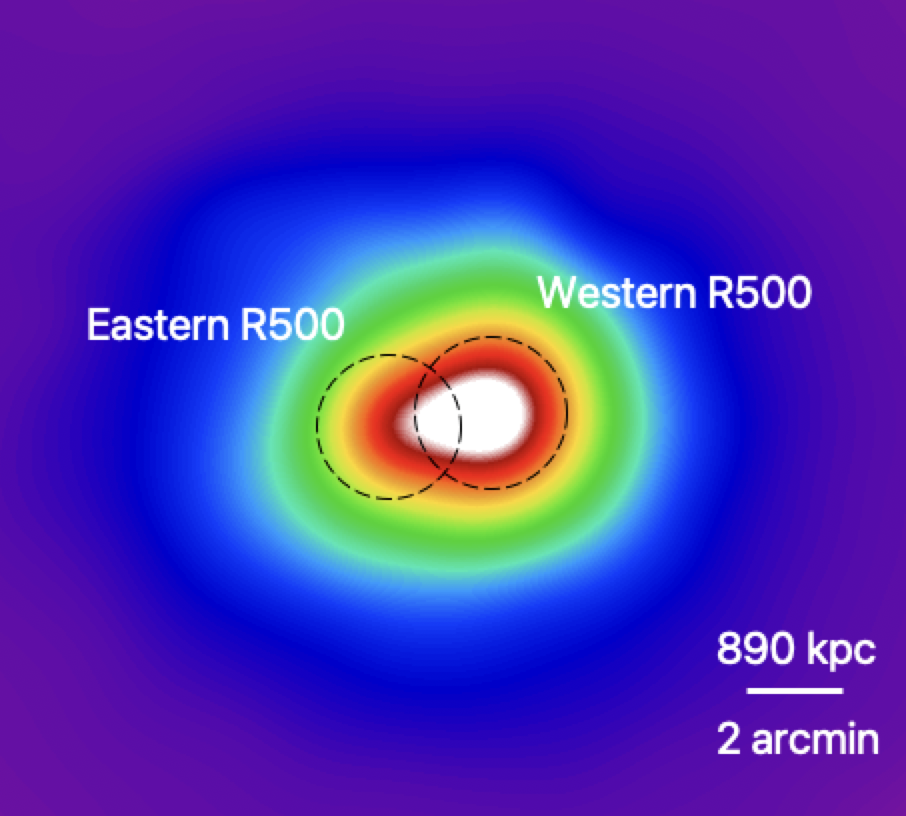}
               \caption{The SPTpol image of the cluster merger (top) and the expected SPT-SZ Compton$-y$ image based on the derived X-ray $n_e$ and \T\ profiles and the SPT-SZ angular resolution. (bottom). The two images are consistent, i.e., the SZ data are consistent with the X-ray analysis of this work.}
        \label{SPTpol}
\end{figure}

As discussed in Sect. \ref{cluster-description}, the past SZ studies of the merger by \citet{bleem15, bocquet19, hilton21} detected a single cluster with the gas thought to be between the two X-ray peaks. Therefore, we wish to determine if the SZ image is consistent with the X-ray data revealing two merging clusters, or if there is evidence for unknown systematic biases between the SZ and X-ray datasets. To do so, we use the known X-ray properties of the clusters to construct a simulated $y-$image with the angular resolution of SPT-SZ at 150 GHz and compare it with the real image. Specifically, we utilize the known deprojected $n_e(r)$ and $T_{\text{X,3D}}(r)$ to create a 3D $y_{\text{SZ}}$ model of the system, where $y_{\text{SZ}}\propto \int _{\text{los}} n_e(r) T_{\text{X,3D}}(r)$. We consider the best-fit $n_e(r)$ and $T_{\text{X,3D}}(r)$ profiles of the two clusters and the respective values for the gas bridge, to fully account for any possible $y_{\text{SZ}}$ signal. We then integrate along the line of sight to obtain a simulated $y_{\text{SZ}}$ image with perfect resolution. Finally, we convolve the image with a Gaussian kernel of FWHM=$1.1^{\prime}$ to reproduce the expected SPT-SZ image at 150 GHz based on the X-ray data.\footnote{The angular resolution of different SPT-SZ bands slightly varies. Moreover, it is not perfectly Gaussian. However, this test serves as a visual, qualitative comparison between the real and simulated SZ image. Thus, we consider the assumption of a Gaussian beam shape fixed to the 150 GHz size to be sufficient.} The resulted simulated SPT-SZ image and the real, observed image of the system by SPTpol are displayed in Fig. \ref{SPTpol}. The two images are highly consistent. This indicates that the sole reason for the false inference of the gas position by the past studies is the large angular resolution of SZ surveys. 

SZ cluster counts are extensively used for obtaining cosmological constraints \citep[e.g.,][]{planck16, dehaan16, bocquet19,bocquet24b}. As shown in this work, SZ data might confuse individual merging clusters for single objects. This might result in underestimating the total mass of such systems. As an example, for SPT-CLJ2228-5828, the mass estimates from both the SPT-SZ \citep{bleem15,bocquet19} and ACT \citep{hilton21} surveys underestimate the total mass of the cluster merger. Specifically, SPT-SZ returns $\sim 50\%$ and $\sim 30\%$ lower total mass than the X-ray and WL estimates respectively, while ACT returns $\sim 24\%$ and $\sim 11\%$ lower masses respectively. When cosmological forward modeling is utilized, one obtains an observed cluster sample from a simulated mass-selected cluster sample folded with the selection function. The latter should account for beam size effects that might lead to mislabeling merging clusters, such as SPT-CLJ2228-5828. If cosmological simulations accurately recover the fraction of such neighboring cluster systems compared to real data and the selection function efficiently describes what one would observe in the case of such systems, no cosmological bias is expected. This needs to be thoroughly checked when the beam size of the used instrument does not allow for disentangling such structures. On the other hand, if the above conditions are not met, then some propagated bias to the constrained cosmological parameters might be expected. However, such systems are probably rare since when X-ray followup studies are performed \citep[e.g., B19][]{lovisari20}, they do not often reveal such missed mergers. Hence, it is not expected that clusters closely projected on the plane of the sky will significantly alter the cosmological constraints of SZ cluster studies.

\section{Analysis of the gas bridge between the clusters}\label{bridge-sect}

\begin{figure}[hbtp]
               \includegraphics[width=0.49\textwidth]{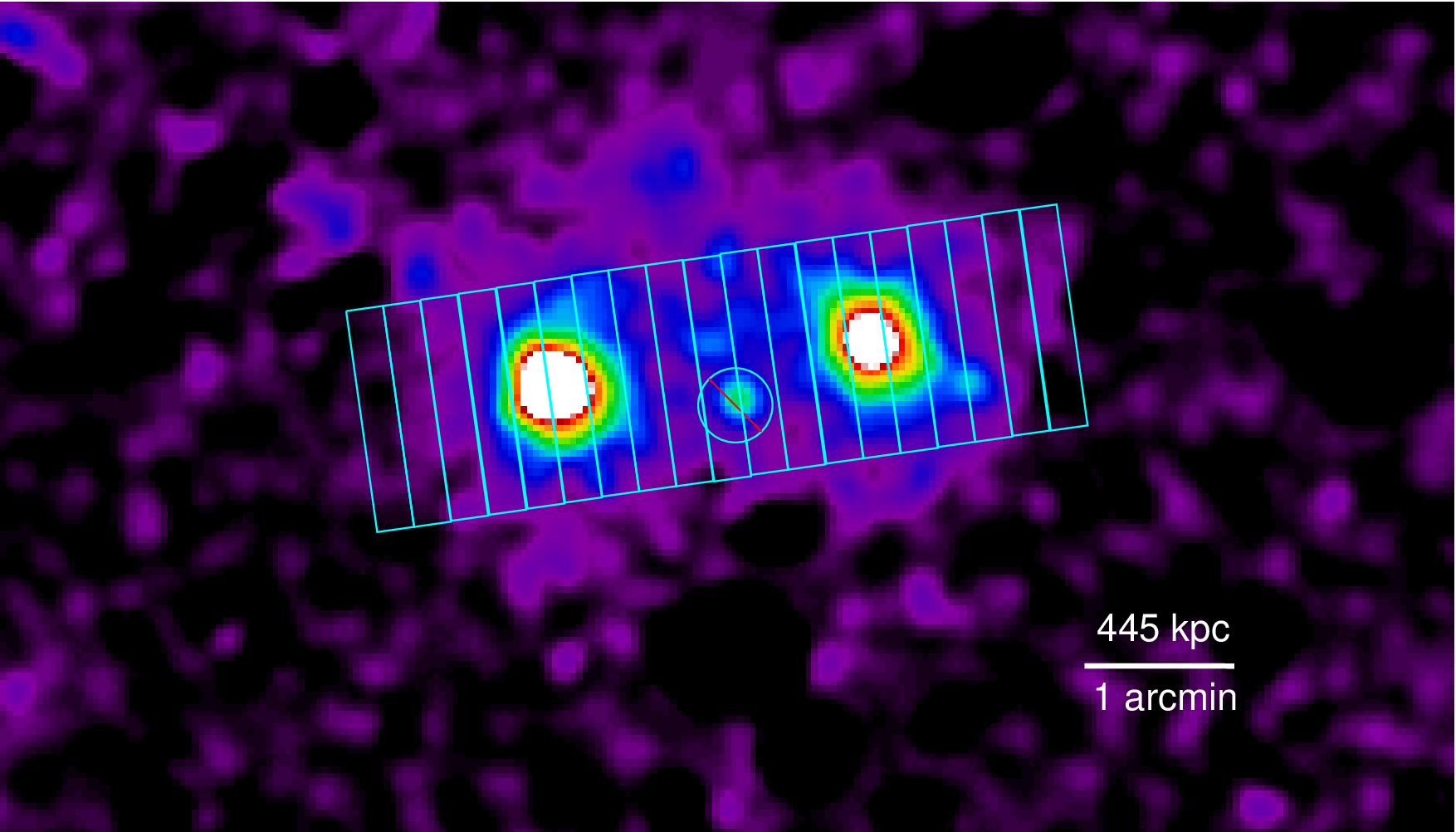}
               \includegraphics[width=0.49\textwidth]{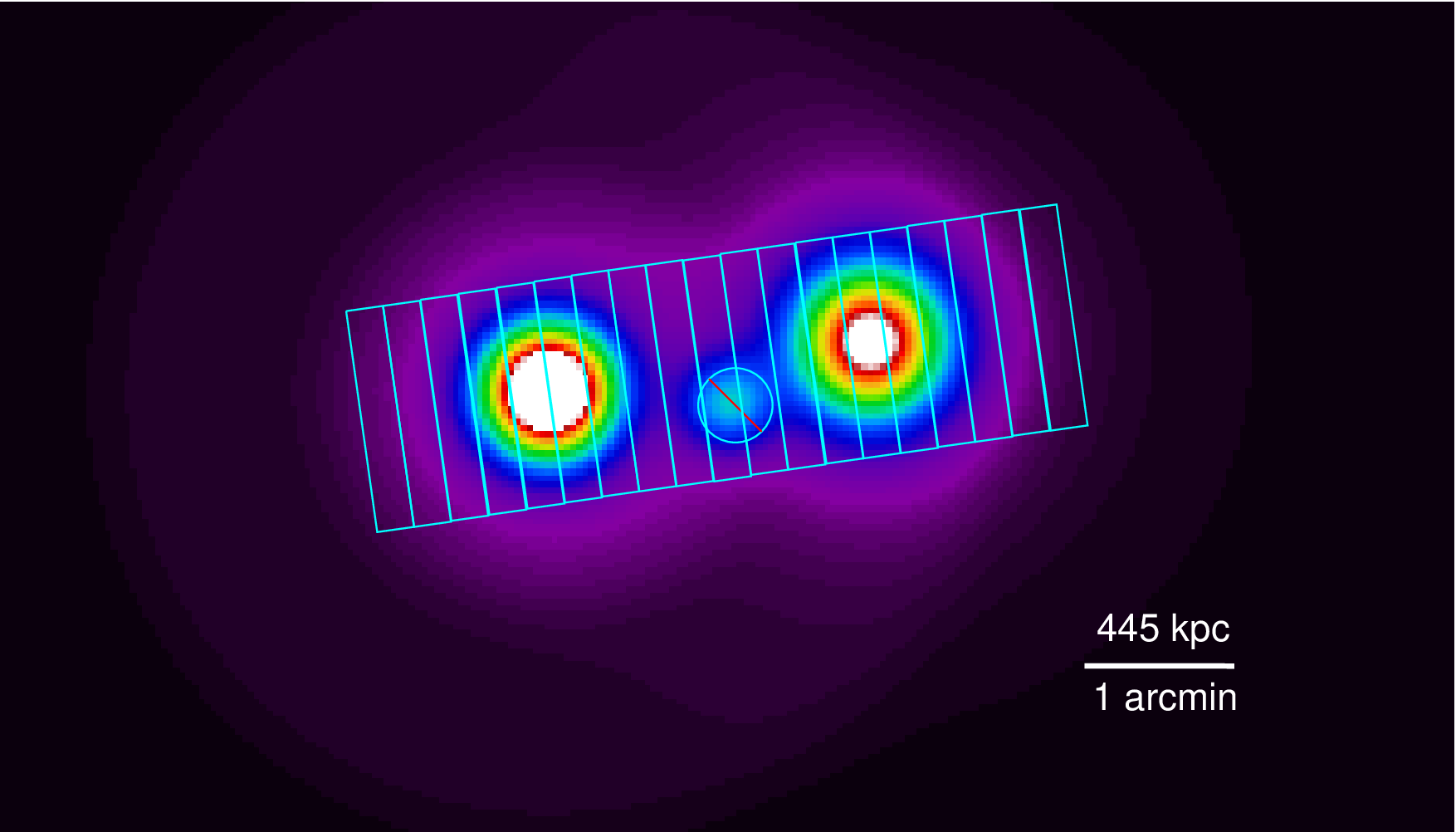}
               \caption{Zoomed-in real (top) and simulated (bottom) image of the system in the $0.5-2$ keV band. The simulated image assumes only the emission of the clusters and no gas bridge. The displayed boxes represent the regions in which the SB and $n_e$ in Fig. \ref{bridge-SB-profile} and \ref{bridge-profiles-rest} were measured.}
        \label{bridge-regions}
\end{figure}

From the X-ray image of the system (Fig. \ref{bridge-regions}), the two clusters seem to interact. However, one cannot immediately distinguish if the image is solely the result of the projection of two clusters at different distances and no interaction takes place. In this case, the apparent emission between the clusters would be caused by the superposition of their outskirt emission. To determine if there is an actual gas bridge one needs to compare the observed emission with the expected emission if all the X-ray signal was originating from the summed cluster outskirt emission. In the following sections we extract the SB, \T, $P$, and $K$ profiles of the region to understand the nature of the existing gas there.

\subsection{Surface brightness profile of the bridge}\label{sect:Sb bridge}
Firstly, we extract the SB profile along the axis of the apparent bridge using 19 box regions, as displayed in the top panel of Fig. \ref{bridge-regions}. Each box has a $90^{\prime \prime}$ height, $15^{\prime \prime}$ width, and an $8^{\circ}$ tilt. There is an AGN located slightly south from the bridge's center at (RA, DEC)$=(337.200^{\circ}, -58.480^{\circ}$) which we mask with a $15^{\prime \prime}$ radius circle (no $z$ information is available for this object and one cannot tell if it is part of the merger). Additionally, we create a simulated image of the merging system inputting the derived SB cluster profiles (Sect. \ref{SB-profiles}) and the CXB level. We assume no excess emission from the bridge on top of the cluster emission. The simulated image was convolved with the \textit{XMM-Newton} EPIC PSF to achieve the same angular resolution as in the real data. We also accurately simulate the AGN emission and mask it as in the real image. The simulated image is displayed in the right panel of Fig. \ref{bridge-regions}. Using the latter, we extract the SB for the boxes between the two clusters.  These values correspond to what one would measure purely due to the sum of the clusters' emission assuming no extra gas. 
%\footnote{The region of interest is $\approx 1'$ away from both cluster cores and as such PSF deconvolution is not necessary for measuring the simulated SB across the bridge.}
\begin{figure*}[h]
\centering
               \includegraphics[width=0.8\textwidth]{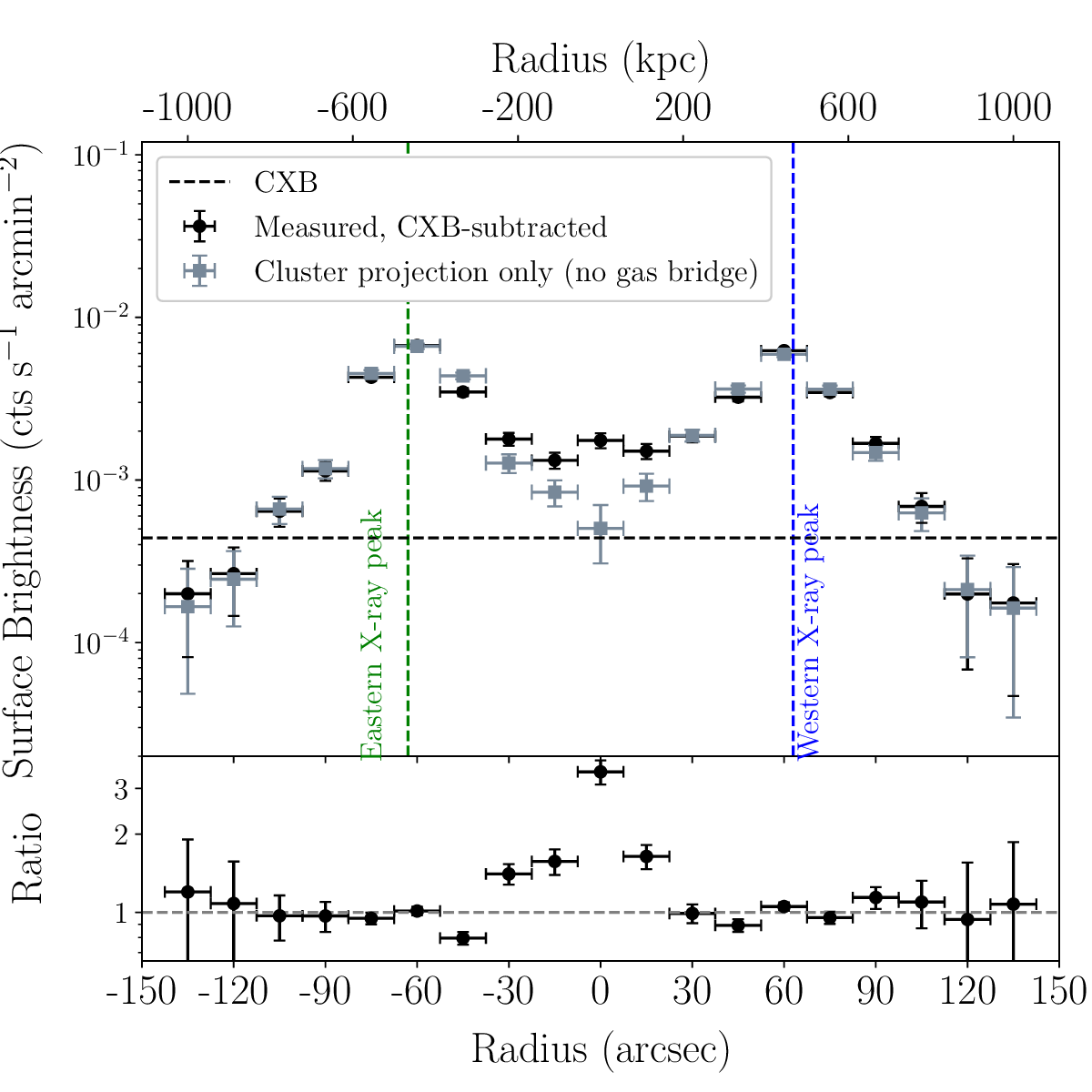}
               \caption{SB profile (CXB-subtracted) across the merging axis for the boxes shown in Fig. \ref{bridge-regions}, measured in the $0.5-2$ keV band. The black points represent the measured SB values. The gray ones represent the expected SB values from cluster projection effects and assuming no excess gas bridge. The position of the Eastern and Western X-ray peaks are displayed by the dashed green and blue vertical lines respectively. The horizontal black dashed line shows the CXB level. The SB ratio between the measured values and the expected values if there was no gas bridge is shown in the bottom panel.}
        \label{bridge-SB-profile}
\end{figure*}

In Fig. \ref{bridge-SB-profile}, we compare the simulated SB values with the true SB across the axis of the bridge. This comparison reveals that there is significant excess emission from the center of the bridge compared to the expected SB from the cluster superposition only. Specifically, there is an SB excess of $(287\pm 49)\%$ in the central box of the bridge at a $5.8\sigma$ level. Moreover, there is a $\approx (57-65)\%$ SB excess in the two neighboring boxes, $15^{\prime \prime}$ (111 kpc) away from the bridge's apparent center. As we move further away from the center of the bridge and toward the cluster cores, there is $\approx 40\%$ excess emission toward the Eastern cluster $30^{\prime \prime}$ (222 kpc) away from the center of the bridge while there is no apparent excess toward the Western cluster at the same distance from the bridge's center. These results strongly demonstrate the existence of a gas bridge between the two clusters with significant X-ray emission. 

The formation of this bridge could follow two scenarios. Firstly, a gas filament could originally connect the two clusters and as they approached each other due to their mutual gravitational attraction, some of the filamentary gas compressed to form the highly emitting gas bridge. If the compression is substantial, weak axial shock fronts could also arise, heating the filamentary gas, similar to \citet{sarkar22}. This scenario would not require the true $R_{500}$ of the clusters to overlap. The second case is that the material of the two clusters has started to collide due to the cluster merging, possibly forming gas clumps and stronger shocks heating the gas. The latter result in significantly higher emission compared to the cluster outskirts. In that case, the cluster material could additionally mix with the gas from the presumed filament that originally connected the two systems. The two scenarios are not easy to distinguish given the observational limitations. Thus, we study additional X-ray properties of the bridge to further understand the exact physical process that formed it.

\subsection{Temperature profile of the bridge}\label{sect:bridge-T}
To extract the \T\ profile across the bridge axis, we need more photons per box than for the SB profiles. Hence, we consider boxes with $31.5^{\prime \prime}$ width and $90^{\prime \prime}$ height, which results in three equally sized, independent \T\ bins between the two cluster centers. We cannot accurately predict the expected \T\ if there was no gas bridge, therefore any such comparison is approximate. The \T\ profile is displayed in the top left panel of Fig. \ref{bridge-profiles-rest}. Across the gas bridge, \T\ shows a significant increase instead of gradually decreasing as one would expect if the X-ray signal was coming only from the cluster outskirts. Specifically, at the bridge's center, we measure \T$=5.53_{-1.09}^{+1.47}$ keV instead of the expected \T$\approx 1.5-2$ keV. The metallicity cannot be constrained and results in $Z=0.29_{-0.29}^{+0.94}\ Z_{\odot}$. Therefore, for the bridge box bins we fix it to the typical value of $Z=0.20\ Z_{\odot}$, as derived by \citet{lovisari19} for cluster outskirts. \footnote{For comparison, we also used $Z=0.10\ Z_{\odot}$ for the bridge box bins but we did not detect significant changes. All new \T\ results fluctuated within $\leq0.5\sigma$ from the default \T\ results using $Z=0.20\ Z_{\odot}$.}.

The position of the \T\ jump coincides with the SB jump in Fig. \ref{bridge-SB-profile} and it demonstrates that the gas there has been heated by a physical mechanism. The most likely scenario is that the hot ICM of the two clusters and the gas bridge has started colliding and compressing, giving rise to a possible shock front, $\approx 475$ kpc away from each cluster's core (assuming the two clusters lie at the exact same cosmic distance). Moreover, this \T\ jump strongly disfavors the first scenario discussed in Sect. \ref{sect:Sb bridge} where the gas bridge was purely coming from a pre-existing filament and the true radii of the clusters did not interact. Such high \T\ is not expected without the interaction of the already-hot ICM of the two systems. Thus, one concludes that the true distance of the clusters should be smaller than the sum of their $R_{500}$. Considering their angular distance, this implies that their radial distance should be $\lesssim 1$ Mpc. As a result, one can safely assume that the clusters lie at the same cosmic distance and their $\Delta z=0.003$ can be attributed to a relative $520$ km s$^{-1}$ peculiar velocity. Finally, the bin between the bridge center and the Eastern cluster also shows an elevated \T\ compared to the symmetrical bin opposite of the Eastern cluster's center. This suggests that any gas interaction and shock front has likely started to extend in that region.

\subsection{Gas density profile of the bridge}
The $n_e$ values for the boxes can be derived from the spectral fits for each region, namely the normalization value of the \texttt{apec} model that describes the thermal emission of the clusters or the bridge. However, this does not allow us to disentangle the $n_e$ coming from the cluster outskirts and the $n_e$ coming from the gas bridge overdensity not associated with the clusters. Consequently, we use a novel approach to estimate the extra $n_e$ originating from the bridge only for each box. Our approach is based on the already extracted $n_e$ profiles from Sect. \ref{sect-ne-profile}. For the following procedure, we assume both clusters to be at the same distance ($z=0.77$) as discussed in the previous section.

We create a 3D model of each cluster using the known cluster $n_e$ profiles, assuming no gas bridge or shock front. We then find the corresponding cluster volume covered by each projected box and measure the average $n_e$ in the box based on the 3D $n_e$ model. We integrate across the length of each box, which we consider to be from the foreground to the background (compared to each cluster's center) $R_{200}$ limits of each cluster. The chosen limits directly and strongly affect the estimated average $n_e$. The latter drops significantly toward the outskirts and, the narrower the box length selected (i.e., integration limits), the higher the final average $n_e$. Hence, it is important to stress that the box $n_e$ values are partially arbitrary and they are hard to physically interpret, as they cover multiple cluster layers.\footnote{Note that one would have a similar issue with the $n_e$ inferred from the spectroscopic analysis and the \texttt{apec} normalization value.} Therefore, they should not be directly compared with the $n_e$ profiles displayed in Fig. \ref{rest-profiles} that correspond to annuli at a fixed distance. The box $n_e$ would naturally be lower. However, they allow us to compare the expected projected $n_e$ in the bridge region with the observed $n_e$ and determine the gas overdensity and length of the bridge. A similar approach for studying similar pre-merger systems was applied by a few past studies \citep[e.g.,][]{omiya}.

\begin{figure*}[hbtp]
               \includegraphics[width=0.49\textwidth]{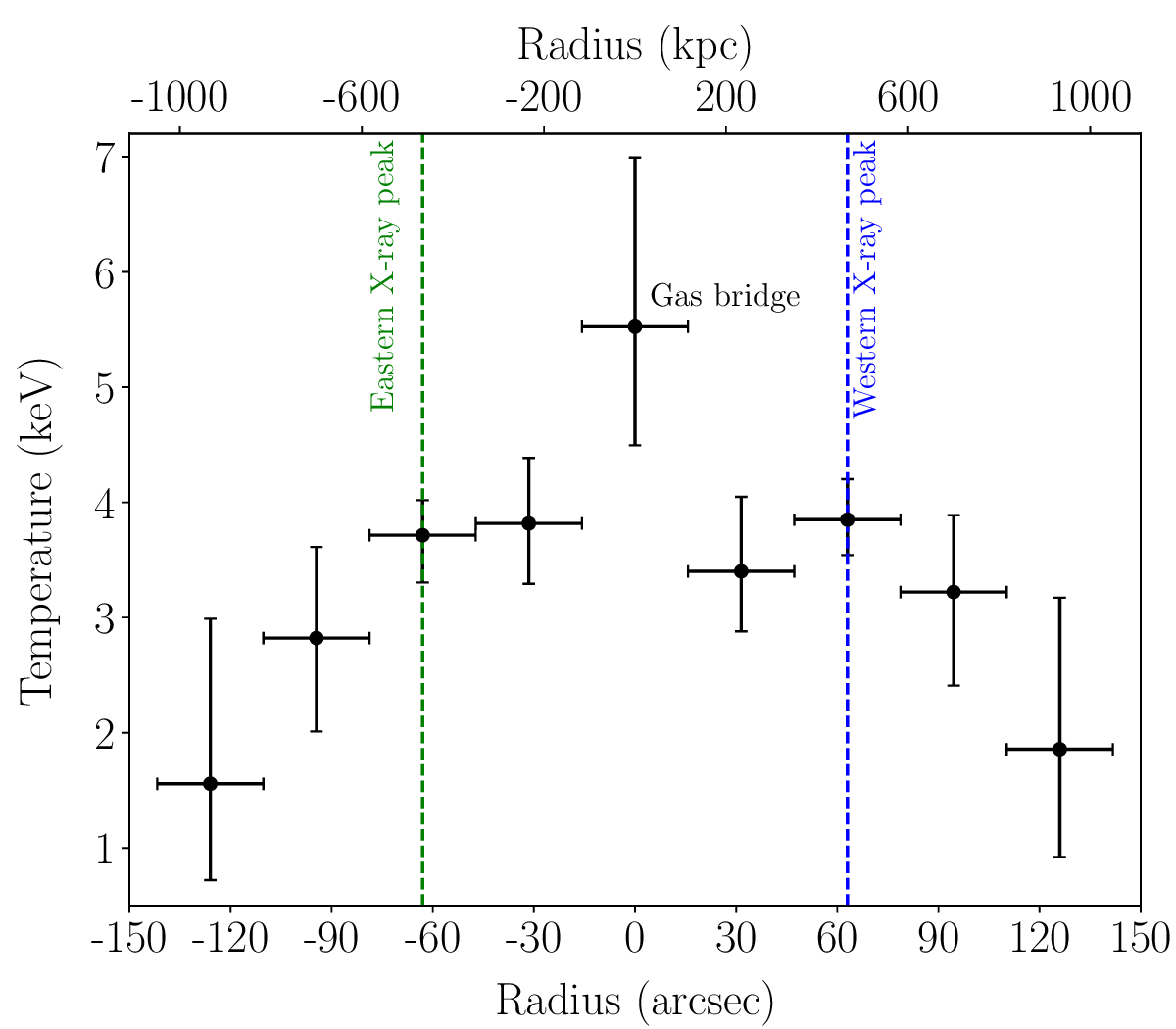}
               \includegraphics[width=0.49\textwidth]{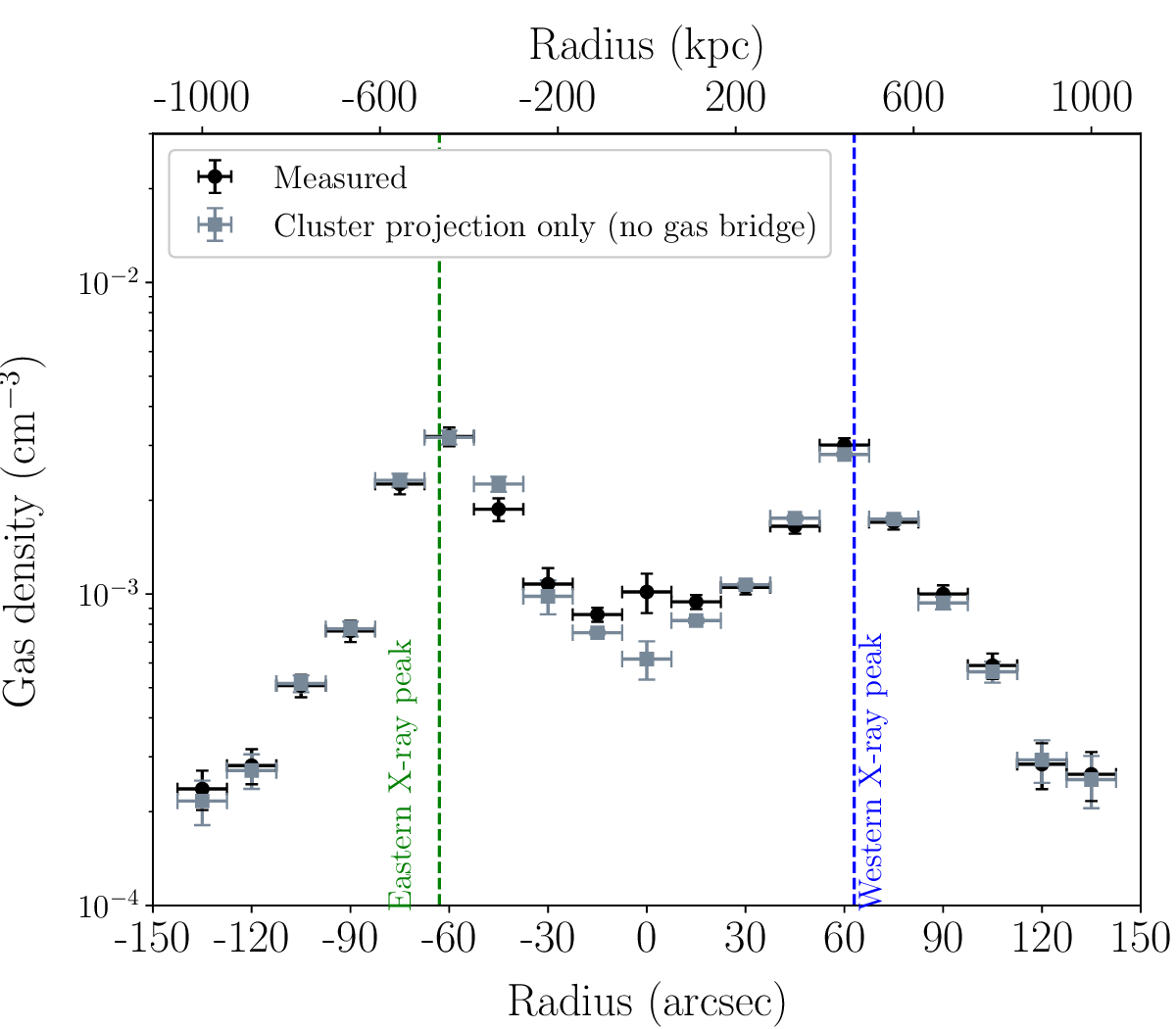}
               \includegraphics[width=0.49\textwidth]{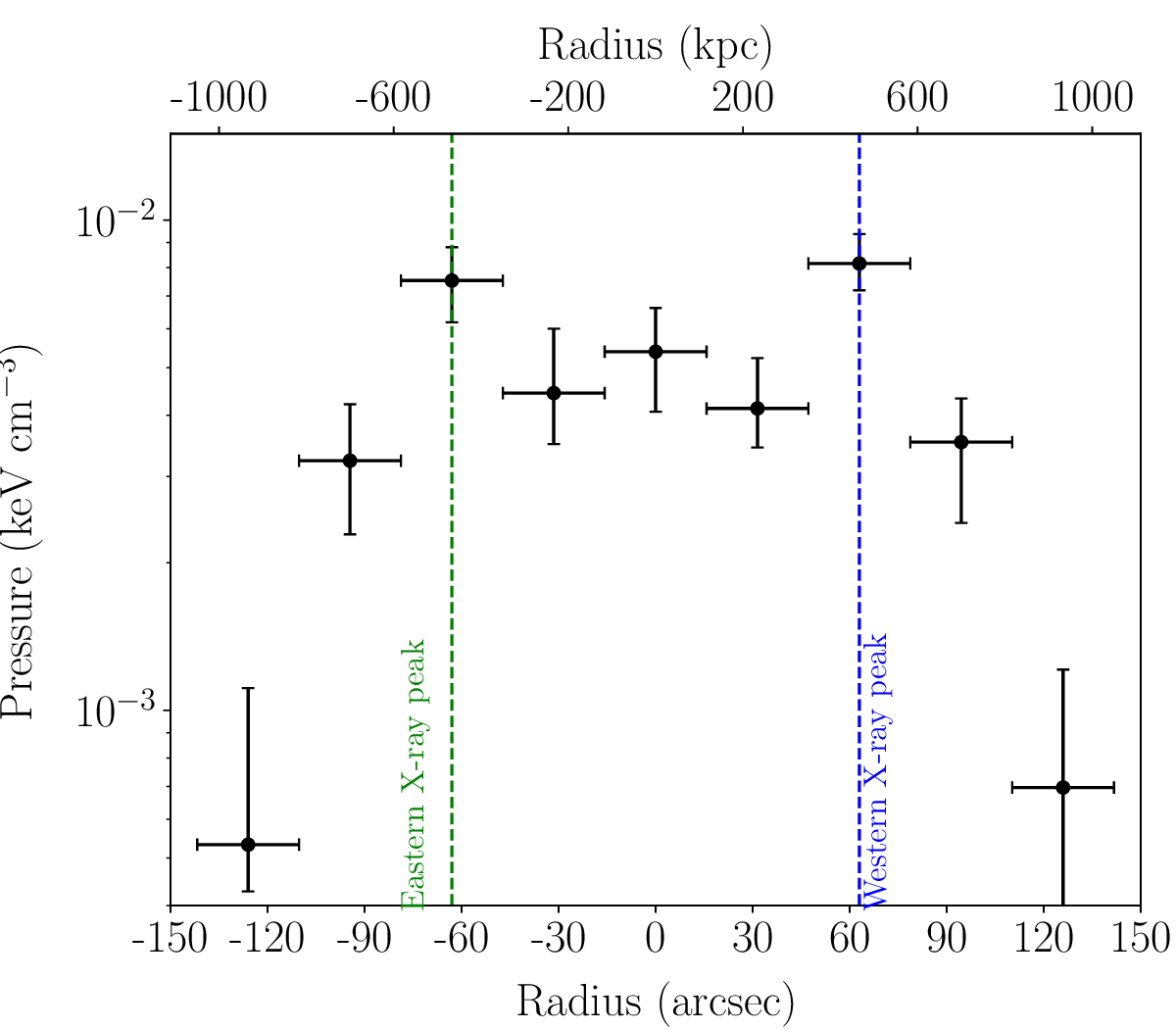}
               \includegraphics[width=0.49\textwidth]{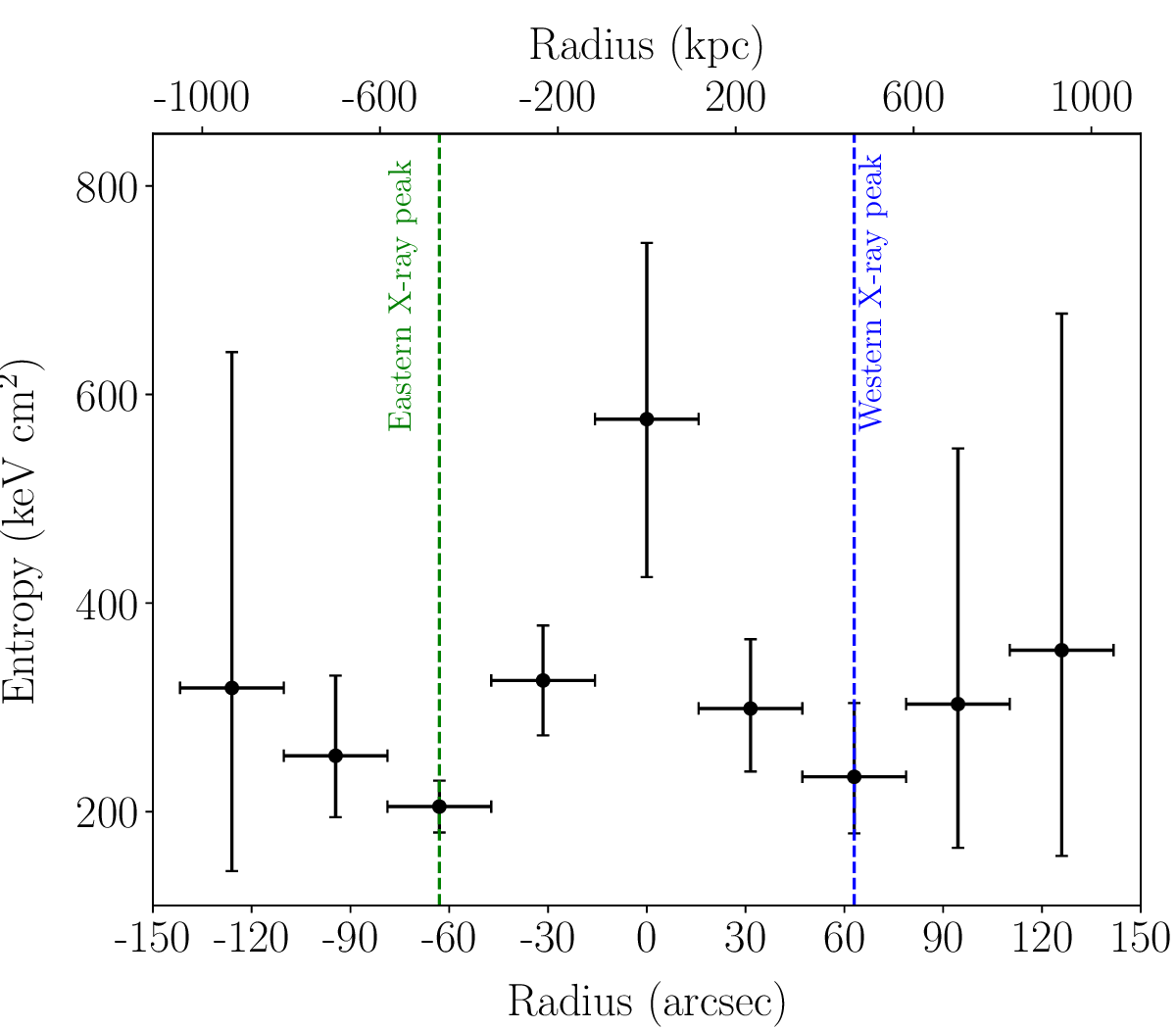}
               \caption{X-ray profiles across the cluster merging axis for the temperature (top left), gas (electron number) density (top right), pressure (bottom left), and entropy (bottom right). The \T, $P$, and $K$ profiles are measured for larger bins than the ones in Fig. \ref{bridge-regions}, with $31.5^{\prime \prime}$ width as explained in Sect. \ref{sect:bridge-T}. The gray points in the $n_e$ profile describe the expected $n_e$ value due to projection effects if no gas bridge was present. All X-ray properties are elevated compared to the case where no gas bridge or shock front exists between the two clusters.}
        \label{bridge-profiles-rest}
\end{figure*}

Due to the coverage from each projected box of several cluster layers and the gas bridge, it is challenging to directly estimate the observed $n_e$ from the SB values. Thus, we follow an empirical approach. We use the SB and $n_e$ values of the 12 independent boxes covering the areas from the cluster X-ray peaks toward their outskirts, opposite to the merging direction. We fit the relation $\text{SB}_i/\text{SB}_j=A\left(n_{e,i}/{n_{e,j}}\right)^B$ where the indexes $i$ and $j$ indicate different box regions. To fit the relation we use all the available box permutations for which $\text{SB}_i/\text{SB}_j<5$, comparable to the $\frac{\text{observed bridge SB}}{\text{predicted bridge SB}}=3.87$ ratio from Sect. \ref{sect:Sb bridge} and avoiding more extreme SB differences. From the obtained best-fit relation, we interpolate the true $n_e$ of the gas bridge boxes by considering the $\frac{\text{observed bridge SB}}{\text{predicted bridge SB}}$ and the expected $n_e$ per box as determined earlier. We compute the statistical uncertainty of the true $n_e$ estimates through error propagation, considering also the (low) scatter of the fitted SB-$n_e$ relation. To account for any neglected systematics in our methodology we add in quadrature an additional $5\%$ uncertainty in all interpolated $n_e$ values (which is a rather conservative approach). The comparison between the observed and the expected $n_e$ per box is shown in the top right panel of Fig. \ref{bridge-profiles-rest}. 

Similarly to the SB comparison, there is a gas overdensity compared to the expected $n_e$ purely from the cluster outskirts. Specifically, the central bridge region shows a $(64\pm 22)\%\ n_e$ excess at a $2.9\sigma$ level.\footnote{Naively one would expect the statistical significance of the $n_e$ excess to be the same as for the SB excess ($5.8\sigma$), but statistical uncertainties from the SB-$n_e$ scaling relation and the ad-hoc systematic uncertainty we introduced decrease the statistical significance of the $n_e$ excess.} Moreover, there is a gas density jump at the bridge's center compared to the nearby regions, further supporting the existence of a shock front. The density of the extra gas at the center of the bridge, on top of the cluster gas, seems to be $n_e=(3.93\pm 1.36)\times 10^{-4}$ cm$^{-3}$. This corresponds to a baryon overdensity of $\delta_b\approx 320$ which hints at a denser gas distribution than typical bridges between clusters, although this could be an overestimate due to the suspected shock front. The total gas density (bridge$+$cluster gas) is $n_e=(1.01\pm 0.15)\times 10^{-3}$ cm$^{-3}$ which corresponds to $\delta_b\approx 825$. Except for the central part of the bridge, the neighboring box regions also show gas overdensities. The extra $n_e$ at $\approx 111$ kpc ($15^{\prime \prime}$) from the center of the bridge is $n_e\approx 1.06-1.22\times 10^{-4}$ cm$^{-3}$, corresponding to $\delta_b\approx 87-98$. Furthermore, at 222 kpc from the bridge's center and toward the Eastern cluster, the extra $n_e\approx 9.4\times 10^{-5}$ cm$^{-3}$ and $\delta_b\approx 74$. Thus, we conclude that the length of the bridge formed by material that does not come from the cluster outskirts is $\approx 333$ kpc ($45^{\prime \prime}$). Finally, from the estimated $n_e$ we can compute the \Mgas\ of the gas bridge material only. We assume a cylinder shape for the bridge with $1040$ kpc radius (equal to the Eastern cluster's $R_{200}$) and $333$ kpc length (covering the four boxes for which we see a $n_e$ overdensity). We find \Mgas$=(1.37\pm 0.34)\times 10^{12}\ M_{\odot}$. Finally, given the measured \T, we can also estimate \Yx$=(7.57\pm 2.3)\times 10^{12}\ \ M_{\odot}\times$keV.

As a cross-check, we additionally derive the total $n_e$ (clusters$+$bridge) using the spectral results and the \texttt{apec} normalization from the central \T\ bin of the bridge. We assume the same cylinder shape for the bridge as before. For converting the \texttt{apec} normalization to $n_e$ we follow \citep[][see their Eq. 18 for details]{veronica22}. For the 333 kpc of the bridge we obtain a total $n_e=(9.94\pm 1.48)\times 10^{-4}$ cm$^{-3}$, which is fully consistent with the result from our empirical approach above and strongly highlights the robustness of our employed methodology used to derive the extra $n_e$ coming purely from the gas bridge.

\subsection{Pressure and entropy profiles of the bridge}

The $P(r)$ and $K(r)$ profiles of the bridge are displayed in the lower panels of Fig. \ref{bridge-profiles-rest}. As they are both a combination of the \T\ and $n_e$ profiles, they are rather noisy. Since we do not have a predicted \T\ for the case of no gas bridge, there is also no such prediction for $P$ and $K$. Nevertheless, one sees that $P$ shows elevated values across the entire bridge and an increase at the central bridge position compared to what one would expect based on the anti-symmetric $P$ values with respect to the cluster centers. Characteristically, 60$^{\prime \prime}$ away from the cluster centers and opposite to the bridge, we measure $P\sim (6-7)\times 10^{-4}$ keV cm$^{-3}$. Assuming the same $T$ at the same distance toward the bridge, and given that $P=n_e\times T_{\text{X}}$, one would expect to measure $P\sim (1-1.5)\times 10^{-3}$ keV cm$^{-3}$ at the middle of the bridge. However, we measure $P=5.12^{+1.44}_{-1.12}\times 10^{-3}$ keV cm$^{-3}$, which constitutes a substantial $P$ increase at the position of the suspected shock front. Similarly, entropy increases at the bridge's center with $K=583^{+169}_{-152}$ keV cm$^2$, showing larger values than the cluster outskirts opposite to the merging direction as presented in Fig. \ref{rest-profiles}, although the difference is not statistically significant due to the large uncertainties. Even if not much information can be robustly drawn from the $P$ and $K$ profiles due to the complex geometry of the boxed covered area, the results further support the existence of a weak-to-moderate shock front.

\subsection{The shock front and the cluster collision geometry}\label{sect-shock}
The detection of shock fronts in cluster (pre-)mergers is relatively rare due to the many requirements that need to be met for such a shock to be formed and observed. From Fig. \ref{bridge-profiles-rest}, the shock front does not seem to have expanded much further than the central 111 kpc, or $15^{\prime \prime}$, from the gas bridge center.\footnote{We believe that part of the shock is obscured by the masked AGN at (RA, DEC)$=(337.200^{\circ}, -58.480^{\circ}$). Exactly on the north side of the mask and within the same box bin, the shock is assumed to be located. Peaked X-ray emission seen there might be confused for another point source but this scenario is strongly disfavored. Firstly, the spectral fit of the region clearly prefers a thermal emission \texttt{apec} model than a power law. Secondly, there is no optical AGN counterpart there or a bright star.} This indicates that the shock front emerged rather recently at the gas bridge due to the recent collision of cluster material. Consequently, it is challenging to determine its direction of motion. 

To characterize the speed of the shock front and estimate its Mach number $\mathcal{M}$ we use Rankine–Hugoniot jump conditions for the density and temperature jumps \citep[e.g.,][]{Landau, russel10}, which result in
\begin{equation}
   \mathcal{M}_{\text{dens}}=\left[\dfrac{2n_{e,1}/n_{e,2}}{\gamma+1-(\gamma -1)n_{e,1}/n_{e,2}}\right] ^{1/2}
\end{equation}
and
\begin{equation}
   \mathcal{M}_{\text{temp}}=\left[\dfrac{(\gamma+1)^2(T_{\text{X},1}/T_{\text{X},2}-1)}{2\gamma(\gamma -1)}\right]^{1/2}
\end{equation}
respectively. In the above equations, $n_{e,1}$/$T_{\text{X},1}$ and $n_{e,2}$/$T_{\text{X},2}$ are the gas density/temperature of the central box bin (where the shock front is located) and the pre-shock area respectively. Moreover, $\gamma=5/3$ is the adiabatic index for a monatomic gas. As the pre-shock area, we consider the regions covered by the two neighboring box bins to the bin where the $n_e$ and \T\ jumps occur. We find $\mathcal{M}_{\text{dens}}=1.09\pm 0.10$ and $\mathcal{M}_{\text{temp}}=1.31\pm 0.43$, with the two values being consistent within the uncertainties and indicating a shock front velocity $u_{\text{sh}}$, close to the speed of sound $c_{\text{s}}=962\pm 52$ km s$^{-1}$. Specifically, for $\mathcal{M}_{\text{dens}}$ and $\mathcal{M}_{\text{temp}}$ we respectively obtain $u_{\text{sh}}=1050\pm 111$ km s$^{-1}$ and $u_{\text{sh}}=1259\pm 419$ km s$^{-1}$. The Mach number indicates a weak shock front. Assuming the two Mach numbers are independent, their weighted mean is $\mathcal{M}_{\text{comb}}=1.11\pm 0.10$, consintent with being subsonic at $\approx 1\sigma$. The $3\sigma$ upper limit for the Mach number is $\mathcal{M}_{\text{comb}}\leq 1.41$.

Given that the shock front seems to be restricted in the central bin, we assume it has travelled $\lesssim 111$ kpc since it was created. Assuming that the shock's velocity is $\sim 1100$ km s$^{-1}$, the age of the shock is $\lesssim 100$ Myr. Furthermore, one can determine the kinetic energy $E_{\text{kin}}$ that flows through the shock front as \citep[e.g.,][]{Gu19,zhang19,omiya}:

\begin{equation}
E_{\text{kin}}=\frac{1.17}{2}m_{\text{p}}n_{e,\text{pre}}u_{\text{sh}}^3 S \left(1-\frac{n_{e,1}^2}{n_{e,2}^2}\right)\,,
\end{equation}
where $m_{\text{p}}$ is the proton mass and $S$ is the 2D area of the shock, which we take to be $S=15^{\prime \prime}\times 90^{\prime \prime}=7.41\times 10^4\ $kpc$^2$. Both the $u_{\text{sh}}$ and $S$ estimates have considerable uncertainties, thus the final $E_{\text{kin}}$ estimate is mostly indicative and not a precise constrain. Nevertheless, we find that $E_{\text{kin}}\sim 2.4\times 10^{44}$ ergs/s; this is a slightly below-average amount of energy for a supersonic shock front which is expected given the low Mach number.

Moreover, the velocity of the shock front is representative of the cluster collision velocity on the plane of the sky \citep[e.g.,][]{omiya}, which seems to be of order $\sim 1000-1300$ km s$^{-1}$. As explained before, the emergence of the shock also implies that the two clusters are almost at the same cosmic distance, i.e., separated by $<(R_{500,\text{west}}+R_{500,\text{east}})\approx 1.4$ Mpc. Such cosmic distance difference would correspond to a cosmological $\Delta z_{\text{Hubble}}\approx 4\times 10^{-4}$. Thus, we can safely assume that $\gtrsim 87\%$ of the observed $\Delta z=0.003$ is due to their relative peculiar motion and their collision velocity in the line of sight is $\sim 520$ km s$^{-1}$. The transverse and line of sight approximate collision velocities imply a total collision velocity of $1100-1400$ km s$^{-1}$, with the axis of motion forming a $\sim 22^{\circ}-27^{\circ}$ angle with the plane of the sky.

%\begin{table*}[htbp]
%\centering
%\caption{X-ray and optical properties of Eastern and Western Clusters.}
%\label{X-ray-table-bridge}
%\begin{tabular}{lcc}
%\hline
%\hline
%Central gas bridge property & Observed & Expected \\
%\hline
%SB & 0.771 & 0.768 \\
%$n_e$ & $0.771^{+0.016}_{-0.012}$ & $0.756^{+0.019}_{-0.019}$ \\
%$T_{\text{X}}$ (keV) & $3.638^{+0.429}_{-0.351}$ & $3.841^{+0.355}_{-0.351}$ \\
%$Z$ ($Z_{\odot}$) & $ 0.29^{+0.94}_{-0.29}$ & - \\
%$M_{\text{gas}}$ ($10^{13}\ M_{\odot}$) & $1.91\pm 0.14$ & $2.19\pm 0.16$ \\
%$Y_{\text{X}}$ ($10^{13}\ M_{\odot}\times$keV) & $6.95^{+0.97}_{-0.84}$ & $8.41^{+1.00}_{-0.98}$\\
%$L$ (kpc) & $688\pm 22$ & $724\pm 22$  \\
%$P$ ($10^{-3}$ keV/cm$^3$) & $4.69\pm 0.31$ & $5.16\pm 0.30$ \\
%$K_{500}$ (keV\ cm$^2$) & $458\pm 31$ & $506\pm 30$ \\
%\hline
%\end{tabular}
%\end{table*}

%\begin{figure*}[hbtp]
 %              \includegraphics[width=0.49\textwidth]{Screenshot 2023-09-10 at 15.34.06.png}
  %             \caption{.}
   %     \label{redshifts}
%\end{figure*}
%\subsection{Potential length of gas bridge}

\section{Summary}\label{sect:summary}
Pre-merging pairs of galaxy clusters are of special interest as they can provide valuable insights into the early phases of cluster interaction, the physical processes taking place during the colliding of the hot ionized gas, formed gas bridges, and the impact of the initial merging on the behavior of the individual clusters. In this work, we performed an in-depth X-ray analysis of SPT-CLJ2228-5828, a distant ($z\approx 0.77$) pair of clusters that have started colliding. From past SZ and optical analyses, SPT-CLJ2228-5828 was %thought
considered to be a candidate for a dissociative post-merger where the gas was separated from the two distinctive galaxy overdensities and WL mass peaks. Our work revealed that this is a pre-merging double cluster system instead, with the radii of the two clusters overlapping and the gas of their outskirts interacting. Our work comprises of two parts; the study of the undisturbed sides of the two galaxy clusters and the analysis of their interaction region and its gas bridge.

When one focuses on the clusters' half-circle regions opposite to the interaction axis, the two separate clusters show highly similar X-ray properties with $T_{\text{X,CE}}\approx 3.2-3.4$ keV, $L_{\text{X,CE}}\approx 0.7\times 10^{44}$ erg s$^{-1}$, \Mgas$\approx (1.8-2.1) \times 10^{13}\ M_{\odot}$, and $Y_{\text{X,CE}}\approx (5.7-7.0)\times 10^{13}\ M_{\odot}\ $keV. Using the \Yx\ mass proxy, we estimate the total cluster mass to be $M_{500,Y_{\text{X}}}=2.10\pm 0.20 \times 10^{14}\ M_{\odot}$ and $M_{500}=2.41\pm 0.22 \times 10^{14}\ M_{\odot}$ for the Eastern and Western cluster respectively. The WL masses of both clusters are also estimated. While the Western cluster is well detected with an estimated $M_{500,{\text{WL}}}$ consistent with the one determined by X-rays, the Eastern cluster is only marginally detected by its WL signal. The total mass of the system as determined by X-rays and WL is respectively $\sim 38\%$ and $\sim 30\%$ higher than the past SPT-SZ results. Moreover, we also extract the SB, $n_e$, \T, $P$, and $K$ profiles of both clusters, which show typical behavior. Interestingly, if one ignores the interacting side of the clusters due to the ongoing merging, both systems consistently follow well-established cluster scaling relations. They also appear to be relaxed, according to their X-ray concentration and central $n_e$ values. All these suggest that interacting clusters may be used as single objects for cosmological and astrophysical studies if one focuses on the cluster sides opposite of the interaction region. Furthermore, by simulating the expected SPT-SZ image of the merger based on its observed X-ray profiles we find that the true SPT-SZ image is consistent with the nature of the merger; thus, the two clusters are detected as one only due to the limited angular resolution of SPT. Finally, assuming the two clusters have the exact same cosmic distance, we conclude that they are colliding with $\sim 1100-1400$ km s$^{-1}$ with a $\sim 22^{\circ}-27^{\circ}$ angle compared to the plane of the sky.

The region between the two merging clusters shows great interest. Firstly, a significant gas overdensity additionally to the cluster outskirts is robustly detected, indicating the existence of a separate gas bridge between the two clusters with $\approx 333$ kpc length. We compute its gas mass to be \Mgas$=(1.37\pm 0.34)\times 10^{12}\ M_{\odot}$. The baryon overdensity of the bridge (excluding the expected gas from the clusters) is $\delta_b\approx 320$ at its center and $\delta_b\sim 75-100$ at its edges. Overall, the studied bridge properties suggest that there was excess gas in-between the clusters before the collision starts, e.g., in the form of a short filament. A weak shock front is found at the center of this gas bridge that shows distinctive gas density and temperature jumps, up to $n_e\approx 10^{-3}$ cm$^{-3}$ and \T$\approx 5.5$ keV respectively. The Mach number is estimated to be $\mathcal{M}\approx 1.1$. The shock front also exhibits a mild pressure and entropy increase compared to the expected value if no shock was present. The age of the shock is estimated to be $\lesssim 100$ Myr and its velocity $u_{\text{sh}}\sim 1100$ km s$^{-1}$. The kinetic energy that flows through the shock is found to be $E_{\text{kin}}\sim 2.4\times 10^{44}$ ergs/s. This is one of the rare detections of shock fronts in interacting, pre-merging clusters. Finally, SPT-CLJ2228-5828 is the only merging system showing a gas bridge and shock front and located at such high-$z$ that has been analyzed using X-ray data.
%and showing a shock front, to be analyzed in X-rays.

\section*{Acknowledgements}

This research is based on observations made with the NASA/ESA Hubble Space Telescope obtained from the Space Telescope Science Institute, which is operated by the Association of Universities for Research in Astronomy, Inc., under NASA contract NAS 5-26555. These observations are associated with GO programme 16488. This work is based on observations obtained at the international Gemini Observatory, a program of NSF NOIRLab, which is managed by the Association of Universities for Research in Astronomy (AURA) under a cooperative agreement with the U.S. National Science Foundation on behalf of the Gemini Observatory partnership: the U.S. National Science Foundation (United States), National Research Council (Canada), Agencia Nacional de Investigaci\'{o}n y Desarrollo (Chile), Ministerio de Ciencia, Tecnolog\'{i}a e Innovaci\'{o}n (Argentina), Minist\'{e}rio da Ci\^{e}ncia, Tecnologia, Inova\c{c}\~{o}es e Comunica\c{c}\~{o}es (Brazil), and Korea Astronomy and Space Science Institute (Republic of Korea). This research makes use of data products derived from observations taken by the 3D-HST Treasury Program
(HST-GO-12177 and HST-GO-12328) with the NASA/ESA Hubble Space Telescope. This study was also based on observations obtained by XMM-Newton, an ESA science mission
with instruments and contributions directly funded by ESA member states and the USA (NASA). These observations were acquired through a successful GO proposal (PI: H. Zohren). K.M. acknowledges support in the form of the X-
ray Oort Fellowship at Leiden Observatory. 
The Bonn and Innsbruck authors acknowledge support from German Federal Ministry for Economic Affairs and Energy (BMWi) provided
through DLR under projects 50OR2002, 50OR2106,  and 50OR2302,
from the German Research Foundation (DFG) under
grant 415537506, and the Austrian Research Promotion
Agency (FFG) and the Federal Ministry of the Republic of Austria for Climate Action, Environment, Mobility, Innovation and Technology (BMK) via grants 899537,
900565, and 911971. Support for Program number GO-16488 was provided through a grant from the STScI under NASA contract NAS5-26555.

%\section{Appendix}

\bibliographystyle{aa} %% aa.bst but adding links and notes to references
%%\raggedright              %% only for adsaa with dvips, not for pdflatex
\bibliography{main}          %% XXX.bib = your Bibtex entries copied from ADS

%\appendix

\end{document}